\documentclass[traditabstract]{aa}

\usepackage{amsmath}
\usepackage{amssymb}
\usepackage{txfonts}
\usepackage{epsfig,graphicx}
\usepackage[colorlinks=true, linkcolor=blue, citecolor=blue, urlcolor=blue]{hyperref}
\usepackage{natbib}
\usepackage{color}
\bibpunct{(}{)}{;}{a}{}{,} 
\begin{document}

\title{The Herschel Virgo Cluster Survey\thanks{Herschel is an ESA space observatory with science instruments provided by European-led Principal Investigator consortia and with important participation from NASA.}}
\subtitle{XV. Planck submillimetre sources in the Virgo Cluster}

\titlerunning{HeViCS XV}

\author{M.~Baes\inst{\ref{inst-UGent}} 
\and 
D.~Herranz\inst{\ref{inst-Cantabria}}
\and
S.~Bianchi\inst{\ref{inst-Arcetri}}
\and
L.~Ciesla\inst{\ref{inst-Crete}}
\and
M.~Clemens\inst{\ref{inst-Padova}}
\and
G.~De~Zotti\inst{\ref{inst-Padova}}
\and
F.~Allaert\inst{\ref{inst-UGent}}
\and
R.~Auld\inst{\ref{inst-Cardiff}}
\and
G.~J.~Bendo\inst{\ref{inst-Manchester}}
\and
M.~Boquien\inst{\ref{inst-Marseille}}
\and
A.~Boselli\inst{\ref{inst-Marseille}}
\and
D.~L.~Clements\inst{\ref{inst-Imperial}}
\and
L.~Cortese\inst{\ref{inst-ESO}}
\and
J.~I.~Davies\inst{\ref{inst-Cardiff}}
\and
I.~De~Looze\inst{\ref{inst-UGent}}
\and
S.~di~Serego~Alighieri\inst{\ref{inst-Arcetri}}
\and
J.~Fritz\inst{\ref{inst-UGent}}
\and
G.~Gentile\inst{\ref{inst-UGent},\ref{inst-VUB}}
\and
J. Gonz\'alez-Nuevo\inst{\ref{inst-Cantabria}}
\and
T.~Hughes\inst{\ref{inst-UGent}}
\and
M.~W.~L.~Smith\inst{\ref{inst-Cardiff}}
\and
J.~Verstappen\inst{\ref{inst-UGent}}
\and
S.~Viaene\inst{\ref{inst-UGent}}
\and
C.~Vlahakis\inst{\ref{inst-ALMA}}
}

\institute{
Sterrenkundig Observatorium, Universiteit Gent, Krijgslaan 281, B-9000 Gent, Belgium\label{inst-UGent}
\and 
Instituto de F\'{\i}sica de Cantabria (CSIC-UC), Avda. los Castros s/n, 39005 Santander, Spain\label{inst-Cantabria}
\and
INAF - Osservatorio Astrofisico di Arcetri, Largo E. Fermi 5, 50125, Florence, Italy\label{inst-Arcetri}
\and
University of Crete, Department of Physics, Heraklion 71003, Greece \label{inst-Crete} 
\and 
INAF-Osservatorio Astronomico di Padova, Vicolo dell'Osservatorio 5, 35122 Padova, Italy\label{inst-Padova}
\and
School of Physics and Astronomy, Cardiff University, Queens Buildings, The Parade, Cardiff CF24 3AA, UK\label{inst-Cardiff}
\and
UK ALMA Regional Centre Node, Jodrell Bank Centre for Astrophysics, School of Physics and Astronomy, University of Manchester, Oxford Road, Manchester M13 9PL, United Kingdom\label{inst-Manchester}
\and
Aix Marseille Universit\'e, CNRS, LAM (Laboratoire d'Astrophysique de Marseille) UMR 7326, 13388 Marseille, France\label{inst-Marseille} 
\and
Imperial College London, Blackett Laboratory, Prince Consort Road, London SW7 2AZ, UK\label{inst-Imperial}
\and
Centre for Astrophysics and Supercomputing, Swinburne University of Technology, Hawthorn, Victoria, 3122, Australia\label{inst-ESO}
\and
Department of Physics and Astrophysics, Vrije Universiteit Brussel, Pleinlaan 2, 1050 Brussels, Belgium\label{inst-VUB}
\and
Joint ALMA Observatory / European Southern Observatory, Alonso de Cordova 3107, Vitacura, Santiago, Chile\label{inst-ALMA}
}

\date{\today}
 
\abstract{
We cross-correlate the Planck Catalogue of Compact Sources (PCCS) with the fully sampled 84 deg$^2$ Herschel Virgo Cluster Survey (HeViCS) fields. We search for and identify the 857 and 545 GHz PCCS sources in the HeViCS fields by studying their FIR/submm and optical counterparts. We find 84 and 48 compact Planck sources in the HeViCS fields at 857 and 545 GHz, respectively. Almost all sources correspond to individual bright Virgo Cluster galaxies. The vast majority of the Planck detected galaxies are late-type spirals, with the Sc class dominating the numbers, while early-type galaxies are virtually absent from the sample, especially at 545 GHz. We compare the HeViCS SPIRE flux densities for the detected galaxies with the four different PCCS flux density estimators and find an excellent correlation with the aperture photometry flux densities, even at the highest flux density levels. We find only seven PCCS sources in the HeViCS fields without a nearby galaxy as obvious counterpart, and conclude that all of these are dominated by Galactic cirrus features or are spurious detections. No Planck sources in the HeViCS fields seem to be associated to high-redshift proto-clusters of dusty galaxies or strongly lensed submm sources. Finally, our study is the first empirical confirmation of the simulation-based estimated completeness of the PCCS, and provides a strong support of the internal PCCS validation procedure.
}

\keywords{galaxies: ISM, submillimeter: galaxies}

\maketitle

\section{Introduction}

Roughly half of all the energy emitted by stars and AGNs in the Universe has been absorbed by dust and re-emitted at far-infrared (FIR) and submillimetre (submm) wavelengths. For many years, this important wavelength regime was one of the least explored windows. Several space missions have been operational in the FIR wavelength region since the mid 1980s, including the Infrared Astronomical Satellite \citep[IRAS,][]{1984ApJ...278L...1N}, the Infrared Space Observatory \citep[ISO,][]{1996A&A...315L..27K}, the Spitzer Space Telescope \citep{2004ApJS..154....1W} and the Akari mission \citep{2007PASJ...59S.369M}. At the same time, the submm window has been explored using bolometer instruments on ground-based telescopes including SCUBA and SCUBA-2 on the James Clerck Maxwell Telescope \citep{1999MNRAS.303..659H, 2006SPIE.6275E..45H} and LABOCA on the Atacama Pathfinder Experiment \citep{2009A&A...497..945S}, as well as balloon experiments such as the  Balloon-borne Large Aperture Submillimeter Telescope \citep[BLAST,][]{2008ApJ...681..400P}. All these missions suffered from relatively poor resolution and/or sensitivity, and left the wavelength region between 200 and 800 $\mu$m virtually unexplored. The simultaneous launch of the Herschel Space Observatory \citep{2010A&A...518L...1P} and the Planck mission \citep{2011A&A...536A...1P} in May 2009 has finally opened the FIR/submm window in earnest.

Herschel was a FIR/submm observatory facility that has offered for the first time imaging and spectroscopic capabilities from space in the wavelength range between 55 and 671~$\mu$m. It contained three instruments onboard: the FIR imager and spectrometer PACS \citep{2010A&A...518L...2P}, the submm imager and spectrometer SPIRE \citep{2010A&A...518L...3G}, and the high-resolution heterodyne spectrometer HIFI \citep{2010A&A...518L...6D}. The PACS camera had three photometric bands, centered at 70, 100 and 160~$\mu$m, the corresponding SPIRE camera operates at three bands centered at 250, 350 and 500 $\mu$m. The angular resolution of the Herschel imaging data varies from 6 arcsec at the PACS 70 $\mu$m band to 36 arcsec at the SPIRE 500 $\mu$m band. The main goal of Herschel was to study the cool interstellar dust and gas in the Universe, with pointed observations of all types of objects, ranging from the Solar System \citep[e.g.,][]{2010A&A...518L.146M, 2011Natur.478..218H} to the deep extragalactic sky \citep[e.g.,][]{2010A&A...518L..11M, 2012MNRAS.424.1614O, 2013ApJ...768...58T}.

Planck is an all sky survey mission that has been simultaneously and continuously scanning the entire sky in nine submm/mm wavebands. It has two instruments onboard: the High Frequency Instrument \citep[HFI,][]{2010A&A...520A...9L} contains bolometers operational at six bands centered at 857, 545, 353, 217, 143 and 100~GHz (corresponding to wavelengths of 350, 550, 850 $\mu$m and 1.38, 2.1 and 3 mm, respectively) and the Low Frequency Instrument \citep[LFI,][]{2010A&A...520A...4B} contains radiometers that cover three bands at 70, 44 and 30 GHz (corresponding to 4.3, 6.8 and 10 mm, respectively). The angular resolution decreases from 4.33~arcmin at the highest frequency to 32.88~arcmin at the lowest frequency. The prime objective of Planck is to measure the spatial anisotropies of the temperature and the polarization of the cosmic microwave background with an unprecedented sensitivity and resolution \citep{2013arXiv1303.5075P, 2013arXiv1303.5076P, 2013arXiv1303.5083P}. In addition, the Planck mission has produced a treasure of valuable information on the submm/mm properties of foreground objects, both in our Milky Way \citep[e.g.,][]{2011A&A...536A..19P, 2011A&A...536A..21P, 2011A&A...536A..22P} and extragalactic objects \citep[e.g.,][]{2011A&A...536A..15P, 2011A&A...536A..16P, 2013MNRAS.433..695C}. A preliminary catalogue of Planck compact sources, the Early Release Compact Source Catalogue (ERCSC) was publicly released in January 2011 and contains more than 15,000 unique sources \citep{2011A&A...536A...7P}. A similar, but more complete catalogue (more than 25,000 sources), the Planck Catalogue of Compact Sources \citep[PCCS,][]{2013arXiv1303.5088P}, was released in March 2013 as part of the major Planck 2013 data and scientific results release \citep{2013arXiv1303.5062P}.

The complementarity and overlap of the Planck and Herschel missions offers a number of advantages. Planck has the advantage of observing the entire sky, and is likely to discover the most rare and extreme submm sources, including strongly lensed high-redshift starburst galaxies and proto-clusters of luminous submm galaxies. With its much better spatial resolution, Herschel has been ideal to follow up these sources, to determine their nature and to characterize their physical properties (the ERCSC was released a few months before Herschel's second in-flight announcement of opportunity for observing time proposals). More generally, Herschel observations of Planck sources are useful to quantify the boosting of Planck flux densities due to noise peaks and confusion, to provide more accurate positions, and to distinguish between Galactic foreground cirrus and genuine extragalactic point sources. Herschel can also take advantage of the absolute calibration of the Planck HFI instrument. Finally, while both missions have a limited overlap in wavelength coverage, they can take advantage of the complementarity beyond this common range. For Herschel sources, Planck flux densities at frequencies of 545 GHz and lower (i.e.\ wavelengths of 550 $\mu$m and beyond) can be useful to investigate a change in the slope of the spectral energy distribution (SED) of galaxies at about 500 $\mu$m that might be related to very cold dust or might originate from dust with a shallow emissivity function. In turn, SPIRE 250~$\mu$m and PACS data are useful to bridge the gap in the SED between the Planck HFI and the 100~$\mu$m band as observed by IRAS.

The complementarity of Herschel and Planck can be exploited in different ways. A first obvious way is to consider a set of galaxies observed by Herschel and complement the Herschel data with Planck observations. This approach has been applied to several samples of nearby galaxies \citep[e.g.,][]{2012A&A...543A.161C, 2012MNRAS.419.3505D, 2013arXiv1303.5083P}, either as a check on the Herschel photometry or to add additional submm data to the SED beyond the SPIRE 500~$\mu$m limit.  The disadvantage of this approach is that the population of galaxies is fixed beforehand (typically selected in the optical), and hence does not allow for a characterization of the Planck population or a discovery of peculiar Planck sources. An alternative approach is to search for and characterize the Planck sources in large-area "blind" extragalactic surveys with Herschel. A first effort along this line is the recent work by \citet{2013A&A...549A..31H}. They cross-correlate the Planck ERCSC catalogue with Herschel observations taken as part of the Herschel Astrophysical Terahertz Large Area Survey \citep[H-ATLAS,][]{2010PASP..122..499E}. H-ATLAS is a Herschel open time key programme that has surveyed about 550 deg$^2$ of extragalactic sky in five FIR/submm bands, using the PACS and SPIRE instruments in parallel mode. \citet{2013A&A...549A..31H} analyzed the compact Planck submm sources in the so-called Phase 1 fields, a set of equatorial fields with a total survey area of 134.5 deg$^2$. They detected 28 compact Planck sources, 16 of which are most probably high-latitude Galactic cirrus features. 10 of the Planck sources correspond to bright, low-redshift galaxies and one source corresponds to a pair of nearby galaxies. Interestingly, one compact Planck source is resolved by Herschel into a condensation of about 15 faint point sources surrounding an unusually bright submm source, which turned out to be a strongly lensed submm galaxy at $z=3.259$ \citep{2012ApJ...753..134F, 2012ApJ...752..152H}.

In this paper, we investigate the Planck compact submm sources in the survey field of another Herschel extragalactic survey, the Herschel Virgo Cluster Survey \citep[HeViCS,][]{2010A&A...518L..48D}. HeViCS is a Herschel open time key program and has conducted a deep survey of four 16 deg$^2$ fields in the Virgo Cluster. The prime aim of HeViCS is a detailed study of the FIR/submm properties of the galaxies in the Virgo Cluster \citep{2010A&A...518L..49C, 2010A&A...518L..51S, 2010A&A...518L..52G, 2010A&A...518L..53B, 2010A&A...518L..54D, 2011A&A...535A..13M, 2012A&A...542A..32C, 2012MNRAS.419.3505D, 2013A&A...552A...8D} and a study of the environmental effects through comparison with nearby galaxies in less dense environments, such as in the HRS \citep{2010PASP..122..261B, 2012A&A...540A..54B, 2012A&A...543A.161C} or KINGFISH \citep{2011PASP..123.1347K, 2012ApJ...745...95D} surveys. However, given that HeViCS has observed a large area of extragalactic sky, it is very suitable for a cross-correlation between Herschel and Planck. Similar to H-ATLAS, HeViCS is a fully-sampled, blind survey and adopts parallel mode observations with both PACS and SPIRE at five wavelengths between 100 and 500 $\mu$m. The main difference between both surveys, apart from the choice of the survey fields and the specific scientific objectives, is the depth of the observations: while H-ATLAS uses two cross-linked scans of every field, HeViCS has performed eight cross-linked scans of every field, with the aim to reach the 250 $\mu$m confusion limit. Due to the time span between the different observations and the corresponding orientation of the survey fields and the spatial offset between the PACS and SPIRE instruments on the sky, HeViCS has observed a total area of 84 deg$^2$, of which 55 deg$^2$ has been observed to the full depth \citep[for details, see][]{2013MNRAS.428.1880A}. Combined with the fact that the Virgo Cluster has been observed extensively at virtually all wavelengths from X-rays to radio waves \citep[e.g.,][]{1994Natur.368..828B, 2004MNRAS.349..922D, 2007MNRAS.379.1599L, 2011A&A...528A.107B, 2012ApJS..200....4F}, the HeViCS fields are ideal for a flux density comparison between Planck and Herschel and a characterisation of the nature of the detected Planck submm sources within the survey field.

The content of this paper is as follows: in Section~2 we search for and characterize the PCCS 857 and 545 GHz sources in the HeViCS survey fields and we search for their Herschel FIR/submm and optical counterparts. In Section~3, we analyze different characteristics of these populations: we compare the various flavors of the flux densities from the PCCS of the Planck detected sources in the HeViCS fields to the corresponding SPIRE flux densities at 350 and 500 $\mu$m, discuss the nature of sources without obvious counterparts, we investigate the completeness and positional accuracy of the PCCS, and we discuss the overdensity of the Virgo Cluster at submm wavelengths. Finally, in Section~4 we present our conclusions and summary. 

\section{Identification and characterization of the sources}

We searched the Planck Catalogue of Compact Sources (PCCS) for all sources within the boundaries of the 84 deg$^2$ extended HeViCS fields. In this section we identify these sources and discuss the global properties of the sample of detected 857 and 545 GHz sources.

 \subsection{857 GHz sources}
 \label{857.sec}
 
\begin{table*}[th!]
\centering
\caption{Planck 857 GHz PCCS sources in the extended HeViCS survey area. The first four columns list the PCCS name, the J2000 position of the Planck PCCS source, and the observed PCCS 857 GHz flux density with its associated error (the listed flux density is the standard DETFLUX estimate, see Section~{\ref{FluxComparison.sec}} for details). The following three columns give the name of the optical counterpart (if any), the VCC number, and the galaxy type (from the GOLDMine database). The last column lists the HeViCS SPIRE 350~$\mu$m flux density and its associated error for the sources with an identified counterpart. Details on how this flux density has been obtained can be found in Section~{\ref{FluxComparison.sec}}.}
\label{857.tab}
\begin{tabular}{cccrcccr}
\hline\hline
PCCS source & $\alpha_{\text{J2000}}$ & $\delta_{\text{J2000}}$ & PCCS 857 GHz & 
counterpart & VCC & type & SPIRE 350~$\mu$m \\
& (deg) & (deg) & (Jy)\hspace{2em} & & & & (Jy)\hspace{2em} \\ \hline 
PCCS1 857 G261.44+74.24 & 182.12855 & 14.94011 & $0.746\pm0.136$ & $\cdots$ & $\cdots$ & $\cdots$ & $\cdots$\hspace{2.1em} \\
PCCS1 857 G260.38+75.42 & 182.66101 & 16.03927 & $2.111\pm0.147$ & NGC\,4152 & 25 & Sc & $1.996\pm0.149$ \\
PCCS1 857 G266.31+73.24 & 182.68154 & 13.34018 & $0.591\pm0.110$ & IC\,3029 & 27 & Sc & $0.317\pm0.029$ \\
PCCS1 857 G269.69+72.43 & 183.12050 & 12.12778 & $0.980\pm0.154$ & IC\,769 & 58 & Sb & $0.877\pm0.072$ \\
PCCS1 857 G276.80+67.94 & 183.26859 & 07.04378 & $1.770\pm0.158$ & NGC\,4180 & 73 & Sb & $1.582\pm0.114$ \\
PCCS1 857 G268.39+73.70 & 183.44470 & 13.41055 & $3.140\pm0.175$ & NGC\,4189 & 89 & Sc & $2.687\pm0.196$ \\
PCCS1 857 G265.42+74.97 & 183.46055 & 14.91470 & $7.332\pm0.194$ & M98 & 92 & Sb & $11.571\pm0.840$ \\
PCCS1 857 G268.93+73.52 & 183.48450 & 13.17543 & $1.610\pm0.144$ & NGC\,4193 & 97 & Sc & $1.592\pm0.125$ \\
PCCS1 857 G278.93+66.94 & 183.64987 & 05.81024 & $1.749\pm0.129$ & NGC\,4197 & 120 & Scd & $1.574\pm0.118$ \\
PCCS1 857 G270.21+73.54 & 183.81944 & 13.01505 & $1.942\pm0.151$ & NGC\,4206 & 145 & Sc & $2.205\pm0.149$ \\
PCCS1 857 G268.89+74.36 & 183.91888 & 13.90904 & $4.512\pm0.168$ & NGC\,4212 & 157 & Sc & $4.847\pm0.346$ \\
PCCS1 857 G270.46+73.72 & 183.97418 & 13.13815 & $8.083\pm0.188$ & NGC\,4216 & 167 & Sb & $9.207\pm0.656$ \\
PCCS1 857 G270.58+73.91 & 184.10005 & 13.29014 & $1.586\pm0.188$ & NGC\,4222 & 187 & Scd & $1.559\pm0.115$ \\
PCCS1 857 G267.17+75.77 & 184.29755 & 15.33868 & $2.645\pm0.129$ & NGC\,4237 & 226 & Sc & $2.919\pm0.210$ \\
PCCS1 857 G282.23+65.16 & 184.30274 & 03.67542 & $0.878\pm0.129$ & NGC\,4234 & 221 & Sc & $0.857\pm0.068$ \\
PCCS1 857 G279.82+68.01 & 184.35336 & 06.68030 & $0.814\pm0.149$ & NGC\,4223 & 234 & Sa & $0.521\pm0.052$ \\
PCCS1 857 G279.69+68.53 & 184.49064 & 07.18604 & $1.055\pm0.124$ & NGC\,4246 & 264 & Sc & $1.000\pm0.079$ \\
PCCS1 857 G270.45+75.20 & 184.71895 & 14.42515 & $20.680\pm0.251$ & M99 & 307 & Sc & $24.229\pm1.720$ \\
PCCS1 857 G282.51+66.97 & 184.98252 & 05.35636 & $3.772\pm0.146$ & NGC\,4273 & 382 & Sc & $3.776\pm0.277$ \\
PCCS1 857 G280.56+69.18 & 185.01716 & 07.68557 & $0.985\pm0.144$ & NGC\,4276 & 393 & Sc & $0.651\pm0.059$ \\
PCCS1 857 G284.34+65.49 & 185.24478 & 03.73035 & $1.156\pm0.140$ & NGC\,4289 & 449 & Sbc & $1.059\pm0.081$ \\
PCCS1 857 G277.07+72.84 & 185.31732 & 11.50735 & $1.937\pm0.153$ & NGC\,4294 & 465 & Sc & $1.939\pm0.144$ \\
PCCS1 857 G277.41+72.86 & 185.41840 & 11.48644 & $1.176\pm0.160$ & NGC\,4299 & 491 & Scd & $1.112\pm0.093$ \\
PCCS1 857 G272.51+75.68 & 185.42477 & 14.59761 & $8.296\pm0.168$ & NGC\,4298/4302 & 483/497 & Sc/Sc& $12.573\pm0.651$ \\
PCCS1 857 G284.35+66.28 & 185.47457 & 04.47981 & $17.872\pm0.226$ & M61 & 508 & Sc & $20.528\pm1.455$\\
PCCS1 857 G280.61+70.62 & 185.53208 & 09.03536 & $2.147\pm0.136$ & NGC\,4307 & 524 & Sbc & $1.757\pm0.127$ \\
PCCS1 857 G282.47+68.84 & 185.55513 & 07.13962 & $0.828\pm0.160$ & NGC\,4309 & 534 & Sa & $0.603\pm0.060$ \\
PCCS1 857 G284.61+66.39 & 185.60479 & 04.55969 & $0.822\pm0.137$ & NGC\,4301 & 552 & Sc & $0.454\pm0.049$ \\
PCCS1 857 G271.42+76.60 & 185.63878 & 15.53121 & $1.516\pm0.168$ & NGC\,4312 & 559 & Sab & $1.532\pm0.113$ \\
PCCS1 857 G277.71+73.24 & 185.65326 & 11.80437 & $1.706\pm0.144$ & NGC\,4313 & 570 & Sab & $1.586\pm0.119$ \\
PCCS1 857 G280.71+70.96 & 185.67829 & 09.34033 & $2.502\pm0.177$ & NGC\,4316 & 576 & Sbc & $1.941\pm0.142$ \\
PCCS1 857 G270.81+76.99 & 185.71130 & 15.93892 & $1.098\pm0.200$ & $\cdots$ & $\cdots$ & $\cdots$ & $\cdots$\hspace{2.1em} \\
PCCS1 857 G271.17+76.89 & 185.73530 & 15.81529 & $17.639\pm0.257$ & M100 & 596 & Sc & $25.968\pm1.840$ \\
PCCS1 857 G284.50+67.11 & 185.77029 & 05.26741 & $0.933\pm0.146$ & NGC\,4324 & 613 & Sa & $0.722\pm0.061$ \\
PCCS1 857 G278.76+72.90 & 185.80843 & 11.36855 & $1.634\pm0.131$ & NGC\,4330 & 630 & Sd & $1.547\pm0.118$ \\
PCCS1 857 G282.91+69.24 & 185.83429 & 07.47657 & $1.941\pm0.177$ & NGC\,4334 & 638 & & $1.736\pm0.127$ \\
PCCS1 857 G283.52+68.77 & 185.90001 & 06.96069 & $2.263\pm0.164$ & NGC\,4343 & 656 & Sb & $1.517\pm0.113$ \\
PCCS1 857 G282.53+70.30 & 186.03782 & 08.52197 & $0.821\pm0.153$ & NGC\,4356 & 713 & Sc & $0.539\pm0.047$ \\
PCCS1 857 G283.94+69.33 & 186.21012 & 07.44889 & $0.963\pm0.155$ & NGC\,4370 & 758 & S0 & $0.734\pm0.057$ \\
PCCS1 857 G286.12+66.93 & 186.33071 & 04.92117 & $1.192\pm0.134$ & NGC\,4378 & 785 & Sa & $1.298\pm0.105$ \\
PCCS1 857 G285.56+67.74 & 186.33387 & 05.75341 & $0.904\pm0.155$ & NGC\,4376 & 787 & Scd & $0.506\pm0.044$ \\
PCCS1 857 G281.96+71.82 & 186.35291 & 10.02117 & $2.201\pm0.156$ & NGC\,4380 & 792 & Sab & $2.001\pm0.149$ \\
PCCS1 857 G284.72+69.17 & 186.43134 & 07.21570 & $2.705\pm0.162$ & UGC\,7513 & 827 & Sc & $2.171\pm0.157$ \\
PCCS1 857 G279.14+74.33 & 186.45118 & 12.66357 & $3.153\pm0.176$ & NGC\,4388 & 836 & Sab & $3.272\pm0.234$ \\
PCCS1 857 G281.83+72.27 & 186.46056 & 10.45480 & $0.892\pm0.154$ & NGC\,4390 & 849 & Sbc & $0.833\pm0.070$ \\
PCCS1 857 G284.56+69.49 & 186.46772 & 07.53835 & $0.860\pm0.162$ & IC\,3322 & 851 & Sc & $0.785\pm0.061$ \\
PCCS1 857 G287.41+65.54 & 186.48927 & 03.44221 & $1.990\pm0.184$ & UGC\,7522 & 859 & Sc & $1.167\pm0.087$ \\
PCCS1 857 G274.37+77.10 & 186.49855 & 15.66868 & $1.950\pm0.170$ & NGC\,4396 & 865 & Sc & $1.823\pm0.135$ \\
PCCS1 857 G278.75+74.77 & 186.51927 & 13.10641 & $5.638\pm0.161$ & NGC\,4402 & 873 & Sc & $5.606\pm0.396$ \\
PCCS1 857 G279.83+74.35 & 186.63556 & 12.60895 & $1.243\pm0.171$ & NGC\,4407 & 912 & Sbc & $1.058\pm0.081$ \\
PCCS1 857 G287.47+66.09 & 186.64485 & 03.96705 & $1.378\pm0.142$ & NGC\,4412 & 921 & Sbc & $1.040\pm0.079$ \\
PCCS1 857 G284.86+69.93 & 186.69124 & 07.92894 & $0.939\pm0.123$ & NGC\,4416 & 938 & Sc & $1.011\pm0.081$ \\
PCCS1 857 G284.05+70.86 & 186.69953 & 08.90471 & $1.128\pm0.147$ & UGC\,7546 & 939 & Sc & $1.115\pm0.108$ \\
PCCS1 857 G276.41+76.63 & 186.72600 & 15.04733 & $4.248\pm0.195$ & NGC\,4419 & 958 & Sa & $3.090\pm0.220$ \\
PCCS1 857 G285.63+69.32 & 186.78597 & 07.27420 & $0.879\pm0.172$ & UGC\,7557 & 975 & Scd & $0.692\pm0.069$ \\
PCCS1 857 G283.86+71.40 & 186.79590 & 09.43363 & $1.308\pm0.147$ & NGC\,4424 & 979 & Sa & $0.931\pm0.073$ \\
PCCS1 857 G286.66+67.97 & 186.79940 & 05.87693 & $0.772\pm0.142$ & NGC\,4423 & 971 & Sd & $0.509\pm0.043$ \\
PCCS1 857 G286.52+68.35 & 186.84390 & 06.25747 & $1.927\pm0.162$ & NGC\,4430 & 1002 & Sc & $1.654\pm0.125$ \\
PCCS1 857 G282.36+73.01 & 186.85696 & 11.11080 & $0.904\pm0.137$ & NGC\,4429 & 1003 & S0/a & $0.672\pm0.052$ \\
\hline\hline
\end{tabular}
\end{table*}

\addtocounter{table}{-1}
\begin{table*}[th!]
\centering
\caption{Continued.}
\begin{tabular}{cccrcccr}
\hline\hline
PCCS source & $\alpha_{\text{J2000}}$ & $\delta_{\text{J2000}}$ & PCCS 857 GHz & 
counterpart & VCC & type & SPIRE 350~$\mu$m \\
& (deg) & (deg) & (Jy)\hspace{2em} & & & & (Jy)\hspace{2em} \\ \hline 
PCCS1 857 G280.30+74.83 & 186.92881 & 13.01208 & $3.006\pm0.153$ & NGC\,4438 & 1043 & Sb & $3.373\pm0.257$ \\
PCCS1 857 G285.11+71.32 & 187.16366 & 09.24766 & $1.104\pm0.138$ & NGC\,4451 & 1118 & Sc & $0.847\pm0.068$ \\
PCCS1 857 G289.14+65.84 & 187.25201 & 03.57333 & $1.763\pm0.135$ & NGC\,4457 & 1145 & Sb & $1.600\pm0.116$ \\
PCCS1 857 G286.11+70.88 & 187.35918 & 08.73663 & $1.198\pm0.150$ & NGC\,4469 & 1190 & Sa & $0.788\pm0.061$ \\
PCCS1 857 G287.55+69.00 & 187.36934 & 06.79160 & $0.778\pm0.143$ & IC\,3414 & 1189 & Sc & $0.367\pm0.036$ \\
PCCS1 857 G286.91+70.02 & 187.39922 & 07.83258 & $1.031\pm0.142$ & NGC\,4470 & 1205 & Sc & $0.977\pm0.076$ \\
PCCS1 857 G289.65+66.56 & 187.60654 & 04.23901 & $1.607\pm0.136$ & NGC\,4480 & 1290 & Sb & $1.394\pm0.104$ \\
PCCS1 857 G283.74+74.48 & 187.69562 & 12.38577 & $1.041\pm0.169$ & M87 & 1316 & E & $0.941\pm0.074$ \\
PCCS1 857 G287.69+70.33 & 187.73649 & 08.07343 & $0.961\pm0.145$ & NGC\,4492 & 1330 & Sa & $0.597\pm0.054$ \\
PCCS1 857 G290.53+66.33 & 187.90427 & 03.93704 & $4.098\pm0.174$ & NGC\,4496 & 1375 & Sc & $3.641\pm0.260$ \\
PCCS1 857 G282.29+76.51 & 187.99116 & 14.43010 & $18.381\pm0.246$ & M88 & 1401 & Sbc & $22.165\pm1.564$ \\
PCCS1 857 G283.78+75.59 & 188.04652 & 13.44229 & $0.775\pm0.141$ & NGC\,4506 & 1419 & Sa & $0.178\pm0.026$ \\
PCCS1 857 G283.51+76.21 & 188.17680 & 14.05463 & $1.165\pm0.142$ & IC\,3476 & 1450 & Sc & $1.085\pm0.092$ \\
PCCS1 857 G289.13+71.03 & 188.36439 & 08.64580 & $2.403\pm0.152$ & NGC\,4519 & 1508 & Sc & $2.268\pm0.170$ \\
PCCS1 857 G288.89+71.57 & 188.40485 & 09.19277 & $1.873\pm0.145$ & NGC\,4522 & 1516 & Sc & $1.434\pm0.107$ \\
PCCS1 857 G290.13+70.13 & 188.50485 & 07.69646 & $3.638\pm0.163$ & NGC\,4526 & 1535 & S0 & $2.857\pm0.207$ \\
PCCS1 857 G286.15+75.39 & 188.57616 & 13.08787 & $1.077\pm0.167$ & NGC\,4531 & 1552 & Sa & $0.590\pm0.051$ \\
PCCS1 857 G290.06+70.64 & 188.58279 & 08.20291 & $8.869\pm0.185$ & NGC\,4535 & 1555 & Sc & $14.235\pm1.024$ \\
PCCS1 857 G291.04+68.93 & 188.58782 & 06.46346 & $3.274\pm0.166$ & NGC\,4532 & 1554 & Sm & $2.893\pm0.206$ \\
PCCS1 857 G290.87+69.64 & 188.65963 & 07.16808 & $0.886\pm0.142$ & IC\,3521 & 1575 & Sm & $0.631\pm0.055$ \\
PCCS1 857 G285.72+76.81 & 188.86411 & 14.47983 & $4.691\pm0.173$ & M91& 1615 & Sb & $6.128\pm0.482$ \\
PCCS1 857 G289.78+73.72 & 189.13379 & 11.23932 & $15.958\pm0.216$ & NGC\,4567/4568 & 1673/1676 & Sc/Sc & $15.196\pm1.281$ \\
PCCS1 857 G288.41+75.62 & 189.19517 &13.17062 & $7.469\pm0.162$ & M90 & 1690 & Sab & $8.169\pm0.581$ \\
PCCS1 857 G287.53+76.64 & 189.23359 & 14.21160 & $3.749\pm0.197$ & NGC\,4571 & 1696 & Sab & $3.597\pm0.275$ \\
PCCS1 857 G290.38+74.34 & 189.42769 & 11.81137 & $6.354\pm0.189$ & M58 & 1727 & Sab & $8.286\pm0.616$ \\
\hline\hline
\end{tabular}
\end{table*}
 
The dominant emission mechanism at 857 GHz (350~$\mu$m) is thermal dust emission. For realistic dust temperatures, 857 GHz is already significantly down the Rayleigh-Jeans tail of the blackbody spectrum, so we primarily expect emission by large amounts of cold dust, rather than warm dust emission linked to star formation \citep[e.g.,][]{2010A&A...518L..65B, 2012MNRAS.419.1833B, 2011AJ....142..111B}. This frequency is still too high for free-free emission or synchrotron emission to start playing a significant role. On the other hand, thanks to the negative K-correction, galaxies at intermediate or high redshift can have the peak of their SED around 857 GHz or even lower frequencies, and star-forming galaxies at $z>0.5$ form a substantial contribution to the galaxy population at the frequency of 857 GHz \citep[e.g.,][]{2010A&A...518L...8C, 2011ApJ...742...24L}. Strongly lensed ultra-luminous infrared galaxies at redshifts $z>1$ could also contribute to the population \citep{2007MNRAS.377.1557N, 2012ApJ...755...46L}. So we expect that the counterparts of the Planck sources at 857 GHz would be dominated by nearby galaxies, with a potential contribution of intermediate to high redshift galaxies (in particular if they tend to cluster within the relatively large Planck beam).

\begin{figure*}
\centering
\includegraphics[width=\textwidth]{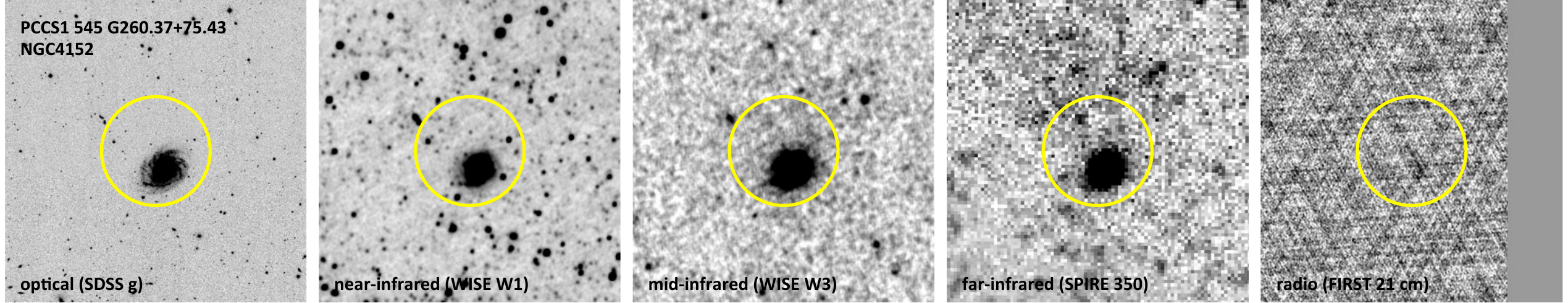}\\[1mm]%
\includegraphics[width=\textwidth]{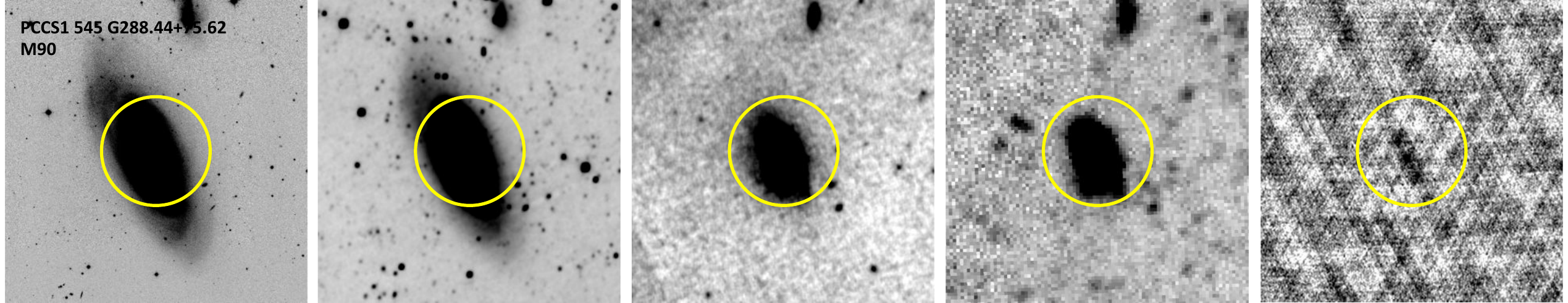}\\[1mm]%
\includegraphics[width=\textwidth]{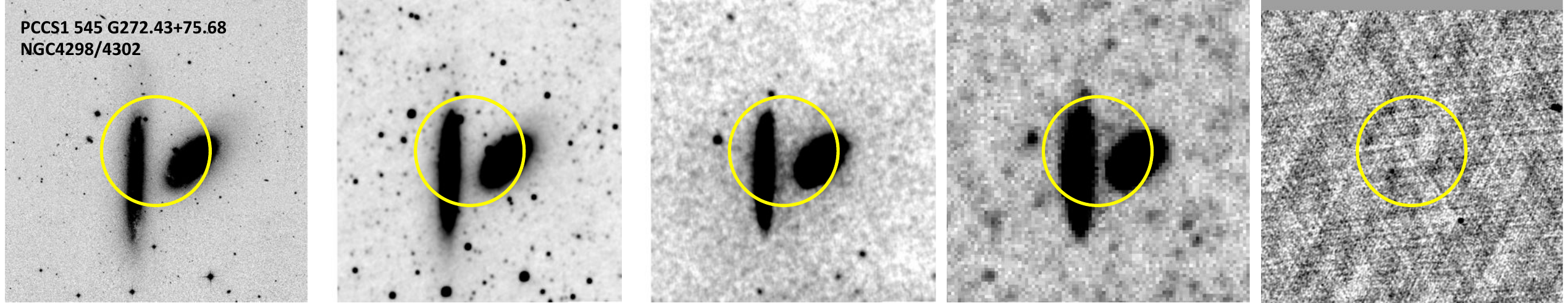}\\[1mm]%
\includegraphics[width=\textwidth]{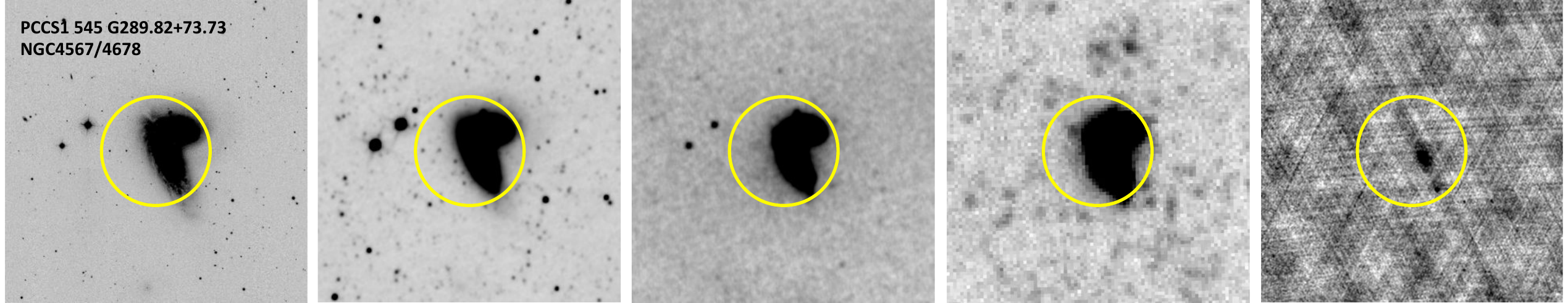}
\caption{Examples of Planck sources with an obvious counterpart at optical and infrared wavelengths. On each row, the different panels represent an optical $g$-band image from SDSS DR9 \citep{2012ApJS..203...21A}, near- and mid-infrared images at 3.4 and 12 $\mu$m from the WISE survey \citep{2010AJ....140.1868W}, the Herschel/SPIRE 350 $\mu$m image from HeViCS, and a 20 cm radio continuum image from the VLA-FIRST radio survey \citep{1995ApJ...450..559B}. Each panel is $12\times12$ arcmin$^2$, and the yellow circle indicates the beam size of the HFI instrument at 857 GHz. The two top rows correspond to sources with individual Virgo Cluster galaxies as counterparts, the two bottom rows correspond to two pairs of Virgo Cluster galaxies.}
\label{DetectedSources.pdf}
\end{figure*}

We used the online Planck Legacy Archive (PLA) tool to select from the PCCS 857 GHz catalogue those sources located in the extended HeViCS survey field. The search resulted in 84 compact Planck sources. In Table~{\ref{857.tab}} we list each of these sources, with the central positions and flux densities as listed in the PCCS. Subsequently, we searched for optical and FIR/submm counterparts for each of these sources, using a simple query in the NASA/IPAC Extragalactic Database (NED) for sources within the 857 GHz beam FWHM and a search in the HeViCS Herschel maps. For 82 of the 84 sources detected at 857 GHz, we could immediately match the Planck source to an optical and Herschel counterpart. Examples of such identifications of Planck sources are shown on Figure~{\ref{DetectedSources.pdf}}.

In 80 cases the counterpart was a single nearby galaxy (top two rows of Figure~{\ref{DetectedSources.pdf}}), and 77 of these belong to the Virgo Cluster. The three remaining ones are the background galaxies IC\,3029, NGC\,4246 and NGC\,4334, at distances of about 98, 56 and 66 Mpc respectively \citep{2005ApJS..160..149S, 2009ApJS..182..474S, 2000ApJ...529..786M}. 

The majority of these 80 nearby galaxies are late-type spiral galaxies, with the Sc class accounting for slightly more than 40\% of all the Planck detected galaxies.  Only three galaxies (NGC\,4370, 4429 and 4526) are classified as lenticular in the Galaxy On Line Database Milano Network \citep[GOLDMine,][]{2003A&A...400..451G}, and the first of these three is sometimes classified as Sa \citep[e.g.,][]{1991rc3..book.....D}. The only elliptical galaxy detected at 857 GHz is M87, the massive radio galaxy at the centre of the Virgo Cluster. Whether or not this galaxy has a substantial interstellar dust reservoir detectable at FIR/submm wavelengths is still a matter of debate, even with the Herschel data at hand \citep{2010A&A...518L..53B, 2010A&A...518L..61B, 2013A&A...552A...8D}. It is clear, however, that the emission at frequencies of 857~GHz and lower is dominated by synchrotron emission that goes as $S_\nu \propto \nu^\alpha$ with a spectral index $\alpha\approx-0.75$ \citep{2007ApJ...655..781S, 2009ApJ...705..356B, 2009ApJ...701.1872C, 2010A&A...518L..53B}.

Two 857 GHz PCCS sources, PCCS1 857 G272.51+75.68 and G289.78+73.72, correspond to a galaxy pair rather than a single bright galaxy (bottom two rows of Figure~{\ref{DetectedSources.pdf}}). The former corresponds to the close pair KPG\,332, formed by the edge-on galaxy NGC\,4302 and the inclined spiral NGC\,4298, with a projected separation of 2.4~arcmin (recall that the Planck's beam size at 857 GHz is 4.33 arcmin). The position of the Planck source is centered between the two galaxies, closer to NGC\,4302, which is the most luminous of both sources at 350~$\mu$m \citep{2012MNRAS.419.3505D, 2013MNRAS.428.1880A}. The latter Planck source PCCS1 857 G289.78+73.72 corresponds to the close pair VV\,219, also known as the Siamese Twins. It consists of the two spiral galaxies NGC\,4567 and 4568, with a separation of 1.3~arcmin. Here the position of the Planck source is close to the centre of NGC\,4568, which is significantly more luminous than its companion at 350~$\mu$m \citep{2012MNRAS.419.3505D, 2013MNRAS.428.1880A}.

The remaining two Planck detections at 857 GHz, PCCS1 857 G261.44+74.24 and G270.81+76.99, do not have any nearby galaxy as optical counterpart.  Their nature will be discussed in Section~{\ref{NoCounterpart.sec}}.

\subsection{545 GHz sources}
 
\begin{table*}[th!]
\centering
\caption{Similar as Table~{\ref{857.tab}}, but for the 545 GHz PCCS sources.}
 \label{545.tab}
 \begin{tabular}{cccrcccr}
\hline\hline
PCCS source & $\alpha_{\text{J2000}}$ & $\delta_{\text{J2000}}$ & PCCS 545 GHz & 
counterpart & VCC & type & SPIRE 500~$\mu$m \\
& (deg) & (deg) & (Jy)\hspace{2em} & & & & (Jy)\hspace{2em} \\ \hline 
PCCS1 545 G260.37+75.43 & 182.66285 & 16.04584 & $0.922\pm0.114$ & NGC\,4152 & 25 & Sc & $0.735\pm0.061$\\
PCCS1 545 G276.77+67.93 & 183.25488 & 07.03960 & $0.533\pm0.104$ & NGC\,4180 & 73 & Sb & $0.562\pm0.044$ \\ 
PCCS1 545 G268.38+73.71 & 183.44879 & 13.41927 & $0.498\pm0.097$ & NGC\,4189 & 89 & Sc & $0.911\pm0.072$ \\
PCCS1 545 G265.43+74.97 & 183.46130 & 14.91524 &$2.640\pm0.107$ & M98 & 92 & Sb & $4.301\pm0.318$ \\
PCCS1 545 G268.93+73.49 & 183.47034 & 13.15472 & $0.548\pm0.111$ & NGC\,4193 & 97 & Sc & $0.646\pm0.060$ \\
PCCS1 545 G278.92+66.95 & 183.64901 & 05.81566 & $0.654\pm0.111$ & NGC\,4197 & 120 & Scd & $0.670\pm0.054$ \\
PCCS1 545 G278.07+68.11 & 183.77733 & 07.01443 & $0.493\pm0.094$ & $\cdots$ & $\cdots$ & $\cdots$ & $\cdots$\hspace{2.1em} \\
PCCS1 545 G270.22+73.54 & 183.81908 & 13.01133 & $0.838\pm0.103$ & NGC\,4206 & 145 & Sc & $0.954\pm0.076$ \\
PCCS1 545 G268.88+74.36 & 183.91488 & 13.90592 & $0.148\pm0.106$ & NGC\,4212 & 157 & Sc & $1.650\pm0.123$ \\
PCCS1 545 G270.44+73.74 & 183.97938 & 13.16155 & $2.269\pm0.099$ & NGC\,4216 & 167 & Sb & $3.454\pm0.252$ \\
PCCS1 545 G270.53+73.94 & 184.09985 & 13.31559 & $0.814\pm0.103$ & NGC\,4222 & 187 & Scd & $0.694\pm0.057$ \\
PCCS1 545 G267.22+75.46 & 184.13517 & 15.07150 & $0.559\pm0.105$ & $\cdots$ & $\cdots$ & $\cdots$ & $\cdots$\hspace{2.1em} \\
PCCS1 545 G267.21+75.76 & 184.29853 & 15.32441 & $0.807\pm0.105$ & NGC\,4237 & 226 & Sc & $0.994\pm0.077$ \\
PCCS1 545 G270.36+75.19 & 184.69368 & 14.43079 & $4.538\pm0.116$ & M99 & 307 & Sc & $7.927\pm0.564$ \\
PCCS1 545 G282.53+66.97 & 184.99044 & 05.36161 & $0.942\pm0.109$ & NGC\,4273 & 382 & Sc & $1.368\pm0.108$ \\
PCCS1 545 G277.04+72.84 & 185.30844 & 11.50868 & $0.707\pm0.139$ & NGC\,4294 & 465 & Sc & $0.805\pm0.064$ \\
PCCS1 545 G272.43+75.68 & 185.41151 & 14.61110 & $2.796\pm0.120$ & NGC\,4298/4302 & 483/497 & Sc/Sc & $4.454\pm0.325$\\
PCCS1 545 G284.36+66.28 & 185.47873 & 04.48218 & $4.789\pm0.116$ & M61 & 508 & Sc & $7.050\pm0.502$ \\
PCCS1 545 G280.54+70.63 & 185.50988 & 09.04869 & $0.569\pm0.114$ & NGC\,4307 & 524 & Sbc & $0.623\pm0.050$ \\
PCCS1 545 G271.16+76.88 & 185.72922 & 15.81079 & $4.626\pm0.114$ & M100 & 596 & Sc & $8.825\pm0.638$ \\
PCCS1 545 G278.76+72.88 & 185.80236 & 11.35158 & $0.523\pm0.100$ & NGC\,4330 & 630 & Sd & $0.635\pm0.053$ \\
PCCS1 545 G281.96+71.79 & 186.34195 & 09.99236 & $0.708\pm0.115$ & NGC\,4380 & 792 & Sab & $0.709\pm0.059$ \\
PCCS1 545 G284.73+69.17 & 186.43614 & 07.21787 & $1.125\pm0.129$ & UGC\,7513 & 827 & Sc & $0.908\pm0.071$ \\
PCCS1 545 G279.13+74.34 & 186.45220 & 12.66788 & $0.964\pm0.111$ & NGC\,4388 & 836 & Sab & $1.218\pm0.093$ \\
PCCS1 545 G274.40+77.10 & 186.50366 & 15.65977 & $0.640\pm0.114$ & NGC\,4396 & 865 & Sc & $0.771\pm0.062$ \\
PCCS1 545 G287.46+65.53 & 186.50593 & 03.43199 & $0.664\pm0.133$ & UGC\,7522 & 859 & Sc & $0.526\pm0.045$ \\
PCCS1 545 G278.76+74.77 & 186.52223 & 13.10600 & $1.747\pm0.116$ & NGC\,4402 & 873 & Sc & $1.951\pm0.142$ \\
PCCS1 545 G284.05+70.82 & 186.68484 & 08.86340 & $0.655\pm0.114$ & UGC\,7546 & 939 & Sc & $0.538\pm0.060$\\
PCCS1 545 G276.42+76.65 & 186.73352 & 15.05677 & $0.834\pm0.125$ & NGC\,4419 & 958 & Sa & $1.074\pm0.934$ \\
PCCS1 545 G286.60+68.36 & 186.87344 & 06.25631 & $0.664\pm0.103$ & NGC\,4430 & 1002 & Sc & $0.599\pm0.052$ \\
PCCS1 545 G280.36+74.81 & 186.93803 & 12.98807 & $1.026\pm0.134$ & NGC\,4438 & 1043 & Sb & $1.208\pm0.110$ \\
PCCS1 545 G289.65+66.58 & 187.61114 & 04.25924 & $0.503\pm0.097$ & NGC\,4480 & 1290 & Sb & $0.551\pm0.045$ \\
PCCS1 545 G283.76+74.47 & 187.69775 & 12.37910 & $1.639\pm0.096$ & M87 & 1316 & E & $1.244\pm0.094$ \\
PCCS1 545 G290.54+66.32 & 187.90694 & 03.92769 & $1.329\pm0.119$ & NGC\,4496 & 1375 & Sc & $1.423\pm0.106$ \\
PCCS1 545 G282.29+76.51 & 187.99024 & 14.42921 & $5.887\pm0.120$ & M88 & 1401 & Sbc & $7.886\pm0.558$ \\
PCCS1 545 G287.49+72.83 & 188.27660 & 10.51415 & $0.548\pm0.112$ & $\cdots$ & $\cdots$ & $\cdots$ & $\cdots$\hspace{2.1em} \\
PCCS1 545 G289.12+71.03 & 188.35920 & 08.65074 & $0.877\pm0.109$ & NGC\,4519 & 1508 & Sc & $0.907\pm0.075$ \\
PCCS1 545 G288.90+71.54 & 188.40228 & 09.16123 & $0.537\pm0.090$ & NGC\,4522 & 1516 & Sbc & $0.539\pm0.044$ \\
PCCS1 545 G290.14+70.13 & 188.50820 & 07.69494 & $0.780\pm0.103$ & NGC\,4526 & 1535 & S0 & $0.929\pm0.071$ \\
PCCS1 545 G289.64+71.15 & 188.55205 & 08.73088 & $0.612\pm0.122$ & $\cdots$ & $\cdots$ & $\cdots$ & $\cdots$\hspace{2.1em} \\
PCCS1 545 G291.04+68.92 & 188.58393 & 06.44702 & $0.827\pm0.088$ & NGC\,4532 & 1554 & Sm & $1.099\pm0.081$ \\
PCCS1 545 G290.09+70.63 & 188.58932 & 08.18835 & $2.967\pm0.110$ & NGC\,4535 & 1555 & Sb & $5.387\pm0.408$ \\
PCCS1 545 G285.69+76.83 & 188.86145 & 14.49705 & $1.525\pm0.096$ & M91 & 1615 & Sb & $2.065\pm0.154$ \\
PCCS1 545 G286.36+76.69 & 188.97390 & 14.32142 & $0.568\pm0.115$ & $\cdots$ & $\cdots$ & $\cdots$ & $\cdots$\hspace{2.1em} \\
PCCS1 545 G289.82+73.73 & 189.14824 & 11.24395 & $4.231\pm0.113$ & NGC\,4567/4568 & 1673/1676 & Sc/Sc & $5.197\pm0.447$ \\
PCCS1 545 G288.44+75.62 & 189.20407 & 13.16827 & $2.241\pm0.106$ & M90 & 1690 & Sab & $2.849\pm0.210$ \\
PCCS1 545 G287.49+76.66 & 189.22770 & 14.22749 & $0.947\pm0.130$ & NGC\,4571 & 1696 & Sc & $1.182\pm0.103$ \\
PCCS1 545 G290.37+74.35 & 189.42504 & 11.82011 & $1.887\pm0.100$ & M58 & 1727 & Sab & $2.877\pm0.218$ \\
\hline\hline
\end{tabular}
\end{table*}
 
The dominant emission mechanism at 545 GHz (550~$\mu$m) is still thermal emission by cool, large dust. Due to the steep spectral slope of a modified blackbody at submm wavelengths, $F_\nu \propto \nu^{2+\beta}$ with $\beta\sim1.5-2$ the emissivity index \citep{2012A&A...540A..54B, 2013MNRAS.436.2435S}, the fluxes of nearby dusty galaxies at 545 GHz are typically a factor 6 lower compared to 857 GHz, and we therefore expect a correspondingly lower number of HFI detections corresponding to local dusty galaxies. On the other hand, the negative K-correction is stronger at 545 GHz and the relative contribution of intermediate and high redshift galaxies to the population counts is expected to be larger.

Querying the Planck PCCS catalogue through the PLA tool, we found 48 compact Planck sources within the extended HeViCS fields at 545 GHz. The list of sources can be found in Table~{\ref{545.tab}}. The characterization of the sample is similar as for the 857 GHz sample: the vast majority (43 out of 48) of the Planck sources can be clearly identified with nearby galaxies. All of the 43 sources with clear optical counterparts are in common with the 857 GHz list, including the two galaxy pairs NGC\,4298/4302 and NGC\,4567/4568. All of the 545 GHz detected galaxies belong to the Virgo Cluster, i.e.\ the three background galaxies detected at 857 GHz are not detected anymore at 545 GHz. 

Concerning the type of the galaxies detected, we note that the relative number of late-type galaxies has even increased compared to the 857 GHz detected population: more than half of all the galaxies detected at 545 GHz are of the Sc type. Remarkably fewer early-type spiral galaxies are detected: of the 10 Sa galaxies detected at 857 GHz, only one single system, NGC\,4419, is also detected at 545 GHz. This galaxy at the centre of the Virgo Cluster is classified as a galaxy 'at the end of the Sa sequence' \citep{1994cag..book.....S}, and is characterized by an unusually high ratio between molecular versus atomic gas \citep{1990ApJ...353..460K}. Concerning the early-type galaxy population, we are left with only two galaxies: the synchrotron-dominated elliptical M87, and the lenticular galaxy NGC\,4526, known to host a prominent dust disc \citep{2006ApJS..164..334F, 2006MNRAS.366.1151S, 2009AJ....137.3053Y}.

The PCCS contains five 545 GHz sources in the extended HeViCS fields without an obvious optical counterpart. Their nature will be discussed in Section~{\ref{NoCounterpart.sec}}.
 
\section{Analysis and discussion}

\subsection{Flux density comparison}
\label{FluxComparison.sec}

\begin{figure*}
\centering
\includegraphics[width=0.6\textwidth]{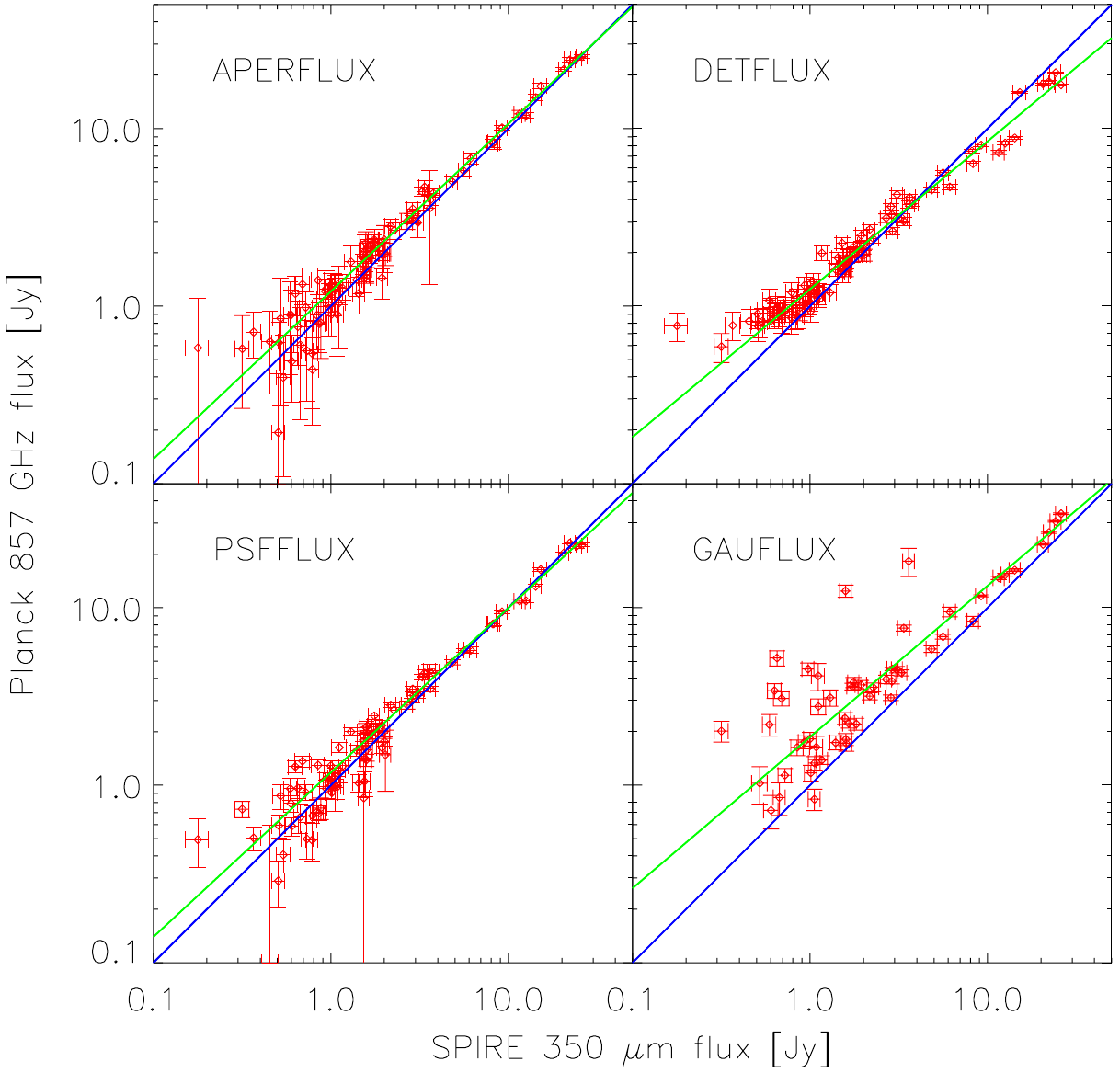}\hspace{4em}
\caption{Comparison between the Herschel SPIRE 350 $\mu$m flux density and the Planck 857 GHz flux density for the 82 PCCS sources with a clear counterpart. The different panels correspond to four different methods used in the PCCS to measure flux densities. The solid green line represents a linear regression fit to the data, the solid blue line corresponds to a one-to-one correlation. The GAUFLUX panel only contains 53 sources as this flux density estimate is not provided for all sources in the PCCS.}
\label{FluxComparison857.pdf}
\end{figure*}
 
\begin{table}[th!]
\centering
\caption{Results of the linear regression fits of the SPIRE versus Planck flux densities. $a$ and $b$ represent the slope and intercept of the linear relation in the form $\log F_{\text{Planck}} = a \log F_{\text{SPIRE}} + b$. The last column gives the reduced $\chi^2$ value of the fit.}
 \begin{tabular}{ccccc}
 \hline\hline
Frequency & Flux density & $a$ & $b$ & $\chi^2_{\text{red}}$ \\ \hline 
857 GHz & APERFLUX & 0.944 & 0.0828 & 0.852 \\
& DETFLUX & 0.833 & 0.0953 & 1.643 \\
& PSFFLUX & 0.927 & 0.0723 & 1.778 \\
& GAUFLUX & 0.852 & 0.271 & 4.314 \\ \hline
545 GHz & APERFLUX & 0.884 & 0.0656 & 1.194 \\
& DETFLUX & 0.792 & 0.0262 & 1.431 \\
& PSFFLUX & 0.864 & 0.0483 & 1.968 \\
& GAUFLUX & 0.815 & 0.165 & 2.950 \\
\hline\hline
\end{tabular}
\label{FluxComparison.tab}
\end{table}

The Planck 857 GHz and SPIRE 350 $\mu$m filters were designed to have virtually the same transmission curve, with nearly identical central wavelength and bandwidth. Therefore, we can immediately compare the PCCS 857 GHz fluxes to the integrated SPIRE 350 $\mu$m fluxes, without significant color corrections. 

SPIRE flux densities, both at 350 and 500 $\mu$m, are based on the results from \citet{2013MNRAS.428.1880A}, who performed automated flux density determination of all Virgo Cluster VCC galaxies in the HeViCS fields. Their flux density determination method consists of several steps, including an advanced determination of the background, an iterative determination of the ideal aperture within which the flux is determined, and an elaborate determination of the uncertainty, taking into account contributions from calibration uncertainty, aperture uncertainty and zero-point uncertainty. The flux densities are presented in their Table~B1. We have applied a number of additional corrections to these values. First, the fluxes were still based on the "old" SPIRE beam sizes, whereas these have been updated since. Secondly, \citet{2013MNRAS.428.1880A} did not contain a correction for extended fluxes, known as the $K_{4E}/K_{4P}$ correction. And finally, the new SPIRE calibration, based on Neptune as a primary calibration source \citep{2013MNRAS.433.3062B}, has been taken into account. The final correction factor comes down to a multiplication by a factor 0.90 and 0.87 at 350 and 500 $\mu$m, respectively. The final integrated flux densities we use for our comparison are listed in the last column of Tables~{\ref{857.tab}} and {\ref{545.tab}} for 350 and 500 $\mu$m respectively.

As mentioned in Section~{\ref{857.sec}}, three galaxies detected at 857 GHz (IC\,3029, NGC4246 and NGC\,4334) do not belong to the Virgo Cluster (even though they all have a number in the VCC). These background galaxies were not considered in Table~B1 of \citet{2013MNRAS.428.1880A}. For these galaxies, we have measured the SPIRE flux densities from the HeViCS maps using exactly the same technique.

Each source in the PCCS has four different measures of the flux density, depending on the source detection algorithm. The detection pipeline flux density (DETFLUX) is directly obtained from the filtered maps and assumes that the sources are pointlike. The three other measures are estimated from the full-sky maps at the positions of the sources. The aperture photometry flux density (APERFLUX) is estimated by integrating the data in a circular aperture centred at the position of the source, with the average FWHM of the effective beam as radius for the aperture. The PSF fit photometry (PSFFLUX) is obtained by fitting a model of the PSF at the position of the source \citep{2011ApJS..193....5M}. Finally, the Gaussian fit photometry flux density (GAUFLUX) is obtained by fitting a Gaussian model, centred at the position of the source, in which the size, shape and background offset are allowed to vary.

\citet{2013arXiv1303.5088P} present a comparison between the different measures for the Planck PCCS 857 GHz flux densities and the SPIRE 350~$\mu$m flux densities for a set of galaxies from four different catalogues\footnote{Comparisons between SPIRE and Planck ERCSC fluxes have also been presented previously for individual galaxy samples, including the HeViCS Bright Galaxy Sample \citep{2012MNRAS.419.3505D}, the Herschel Reference Survey \citep{2012A&A...543A.161C}, and a sample of Planck-selected star forming galaxies \citep{2013MNRAS.429.1309N}.}. They found that, at low flux densities, the smallest dispersion between the SPIRE and Planck flux densities is achieved by the DETFLUX photometry because the filtering process removes structure not associated with compact sources. At the highest levels, corresponding to the most extended galaxies, the flux densities are underestimated by DETFLUX, APERFLUX and PSFFLUX. GAUFLUX accounts for the size of the source and is therefore able to estimate the flux density correctly. The general recommendation from the PCCS is that the appropriate photometry to be used depends on the nature of the source, with GAUFLUX to be used for the brightest and most resolved sources. A similar conclusion was obtained by \citet{2013A&A...549A..31H}.

In Figure~{\ref{FluxComparison857.pdf}} we show the comparison between the Herschel SPIRE 350 $\mu$m flux density and the Planck PCCS 857 GHz flux density, for each of the four different methods used in the PCCS to measure flux densities. The solid blue lines in the different panels correspond to a one-to-one correlation, the solid green lines represent a linear regression fit. These fits were made using the IDL MPFITEXY tool \citep{2010MNRAS.409.1330W}, based on the non-linear least-squares fitting package MPFIT \citep{2009ASPC..411..251M}, and the parameters of these fits can be found in Table~{\ref{FluxComparison.tab}}. 

It is immediately obvious that the GAUFLUX is not an appropriate choice: only the highest flux densities at $S_{350}\gtrsim10$~Jy are recovered fairly reliably, whereas lower flux densities are typically strongly overestimated. Somewhat surprisingly, the DETFLUX flux densities, which showed the smallest dispersion in the comparison of \citet{2013arXiv1303.5088P}, also show systematic deviations from the SPIRE flux densities. At high levels, the flux densities are systematically underestimated, as also found by \citet{2013arXiv1303.5088P}. Also at the lowest flux density levels, however, the DETFLUX measure shows a systematic overestimation compared to the SPIRE flux densities. This might be due to the flux boosting also noted by \citet{2013A&A...549A..31H} for faint ERCSC sources in the H-ATLAS fields. We find the smallest scatter between the Planck and SPIRE flux densities for the APERFLUX flux density estimate. The flux densities agree very well, even up to the highest values where the DETFLUX estimate breaks down. The reduced $\chi^2$ of this correlation is 0.852, which indicates that the error bars on the APERFLUX flux densities might be slightly overestimated. The PSFFLUX flux densities are a close second best, although the scatter at the lowest flux densities is a bit larger. 

Comparing Planck 545 GHz and SPIRE 500 $\mu$m fluxes is less obvious, as the former filter has a significantly longer central wavelength (550 $\mu$m). In order to make a comparison possible, we converted the observed SPIRE 500~$\mu$m fluxes by multiplying it with a conversion factor. The FIR/submm SEDs of nearby galaxies in the Virgo Cluster can be approximated by a modified blackbody, $F_\nu \propto \nu^\beta\,B_\nu(T)$, with a temperature of $T\approx20$~K and $\beta\approx2$ \citep{2012MNRAS.419.3505D, 2013MNRAS.428.1880A}. Taking into account the color corrections of the SPIRE 500 $\mu$m and Planck HFI 545 GHz bands, a color correction of 0.83 has to be applied to the SPIRE flux densities. A similar conversion factor (0.87) was obtained by \citet{2012A&A...543A.161C}, based on a modified blackbody SED with $\beta=1.5$. Only for M87, for which the FIR/submm emission is completely dominated by synchrotron emission rather than by thermal dust emission \citep{2007ApJ...655..781S, 2010A&A...518L..53B}, we applied a different conversion term. For a power-law synchrotron spectrum with $\alpha=-0.75$, we find a conversion term of 1.07.

\begin{figure*}
\centering
\includegraphics[width=0.6\textwidth]{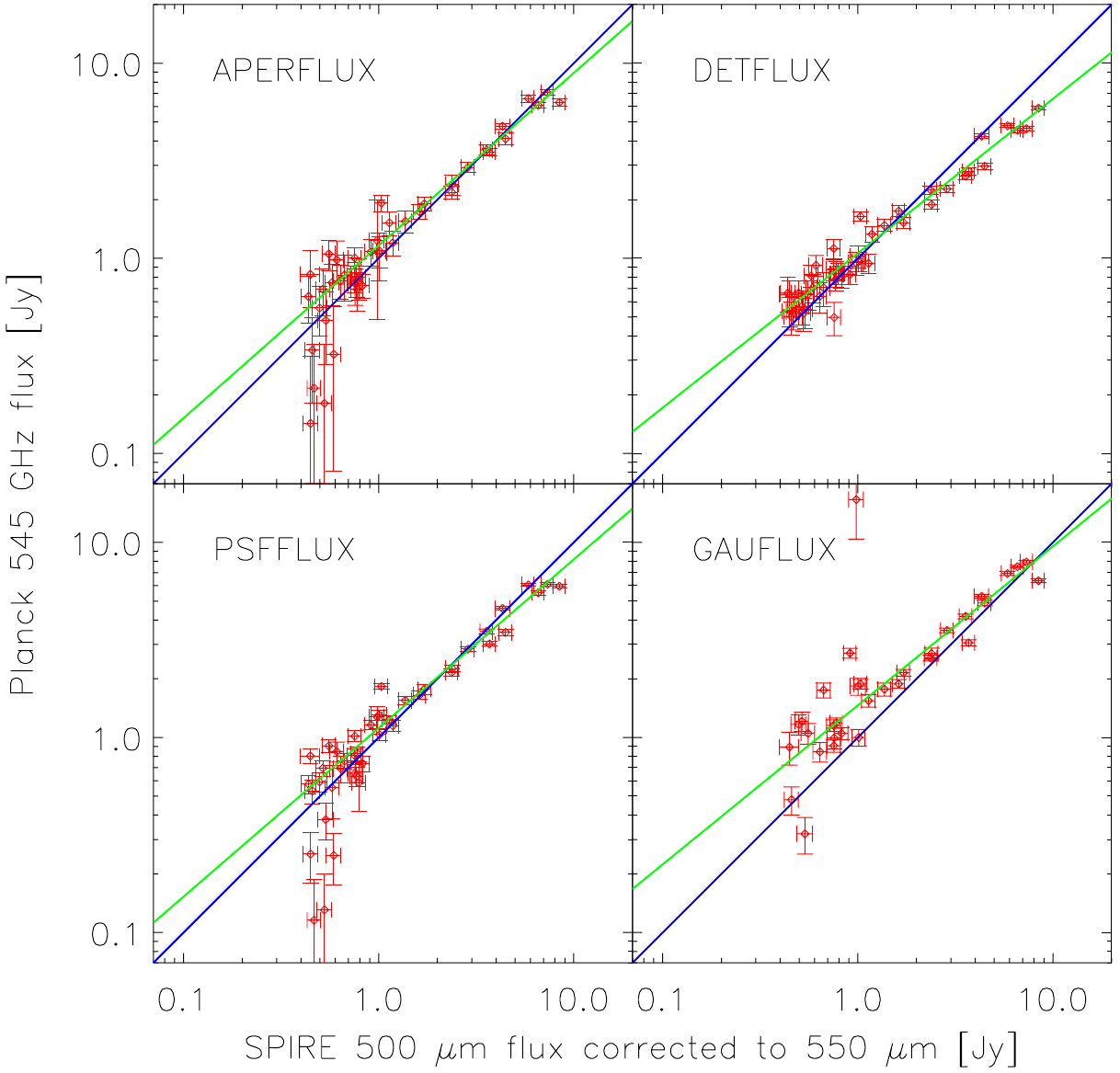}\hspace{4em}
\caption{
Comparison between the Herschel SPIRE 500 $\mu$m flux density and the Planck 545 GHz flux density for the 43 PCCS sources with a clear counterpart. The Herschel flux densities have been converted to correspond to a wavelength of 550 $\mu$m (see text). The different panels correspond to four different methods used in the PCCS to measure flux densities. The solid green line represents a linear regression fit to the data, the solid blue line corresponds to a one-to-one correlation. The GAUFLUX panel only contains 33 sources as this flux density estimate is not provided for all sources in the PCCS.}
\label{FluxComparison545.pdf}
\end{figure*}

The comparison between the observed Planck 550~$\mu$m flux densities and the converted SPIRE 500 $\mu$m flux densities is shown in Figure~{\ref{FluxComparison545.pdf}}, where each panel again corresponds to the different measures for the flux densities in the PCCS. The results of the linear regression fits can be found in Table~{\ref{FluxComparison.tab}}. In general, we find the same behavior as for the comparison of the Planck 857 GHz and SPIRE 350~$\mu$m flux densities. For the highest flux density levels, DETFLUX and PSFFLUX underestimate the true flux densities, whereas GAUFLUX and APERFLUX do not suffer from this effect. Except for the brightest sources, the GAUFLUX estimates are the worst, with an exceptionally spurious flux density estimate more than an order of magnitude off for the otherwise unremarkable Sc galaxy NGC\,4571. In general, the APERFLUX flux density estimates show least dispersion compared to the SPIRE flux densities. In this case, the PSFFLUX flux densities tend to underestimate the true flux densities significantly at the lowest levels. 

\subsection{Modified blackbody fits}
\label{DustProperties.sec}

The FIR/submm traced by Herschel and Planck is dominated by the emission from cold dust. The physical properties of the dust can in principle be determined by fitting SED models to the observed flux densities. While complex multi-component methods have been developed that take into account the entire range from mid-infrared to mm wavelengths \citep[e.g.,][]{2007ApJ...657..810D, 2011A&A...525A.103C, 2011A&A...536A..88G}, the most common approach is a simple modified blackbody fit. This approach enables to determine the most fundamental properties of the dust medium: the mean dust temperature and the amount of dust. 

In the Virgo Cluster, modified blackbody fits have been applied to the FIR/submm SED in several different studies: \citet{2012MNRAS.419.3505D} fitted modified blackbodies to a sample of SPIRE 500 $\mu$m selected galaxies, \citet{2013MNRAS.428.1880A} and \citet{2013arXiv1311.1774D} used them to study the dust characteristics of a samples of optically selected galaxies, and \citet{2010A&A...518L..52G}, \citet{2013A&A...552A...8D} and \citet{2013MNRAS.436.1057D} applied modified blackbody fits to smaller samples of star-forming dwarfs, early-type galaxies and transition-type galaxies, respectively. All of these studies relied on Herschel flux densities. One important question is whether the availability of additional Planck flux densities at 857 and 545 GHz substantially modifies the derived parameters.

In order to test this, we considered all galaxies in the HeViCS fields detected at 545 GHz, and constructed FIR/submm SEDs by combining the Planck HFI with Herschel/SPIRE observations at 250, 350 and 500 $\mu$m, Herschel/PACS data at 100 and 160 $\mu$m. The PACS and SPIRE fluxes and corresponding error bars were taken from \citet{2013MNRAS.428.1880A}, where an automatic flux density measurement was applied to the five HeViCS maps for all galaxies from the Virgo Cluster Catalogue \citep{1985AJ.....90.1681B}. Only M87 was omitted from the sample as its SED is dominated by synchrotron emission rather than cold dust emission. 

We fitted the observed broadband fluxes using a simple modified blackbody function, i.e.\
\begin{equation}
  F_\nu
  =
  \frac{M_{\text{d}}}{D^2}\,\kappa_\nu\,B_\nu(T_{\text{d}})
\end{equation}
with $M_{\text{d}}$ the dust mass, $D$ the distance to the galaxy, $T_{\text{d}}$ the dust temperature and $\kappa_\nu$ the emissivity. We used a common distance of 16.5~Mpc to all galaxies \citep{2007ApJ...655..144M}, and, as customary, we assume a power-law dust emissivity in the FIR/submm wavelength range, i.e.\ $\kappa_\nu \propto \nu^\beta$, where we fixed $\beta$ to 1.8 \citep{2011A&A...536A..25P, 2013MNRAS.436.2435S}, and the zero-point to $\kappa_\nu = 0.192$~m$^2$\,kg$^{-1}$ at 350 $\mu$m \citep{2003ARA&A..41..241D}. Note that both the value of the emissivity index $\beta$ and the absolute calibration of the emissivity are notoriously uncertain; actually, the quoted value of the zeropoint at $\lambda = 350$ $\mu$m has been derived assuming a dust model with $\beta=2$, so it would in principle not be applicable to models with a different value of $\beta$ \citep[for a discussion, see][]{2013A&A...552A..89B}. However, the goal of the present exercise is not to determine absolute dust masses, but rather to compare SED fits with and without Planck data, so the absolute normalization of the emissivity (and the distance to the galaxies) is of less importance in our case. The two remaining free parameters in our fitting routine are the dust mass and the dust temperature. The fits were done by performing a $\chi^2$ minimization using a simple gradient search method. Error bars on the derived parameters were derived using a Monte Carlo bootstrapping method. 

\begin{figure*}
\centering
\includegraphics[width=0.35\textwidth,height=0.35\textwidth]{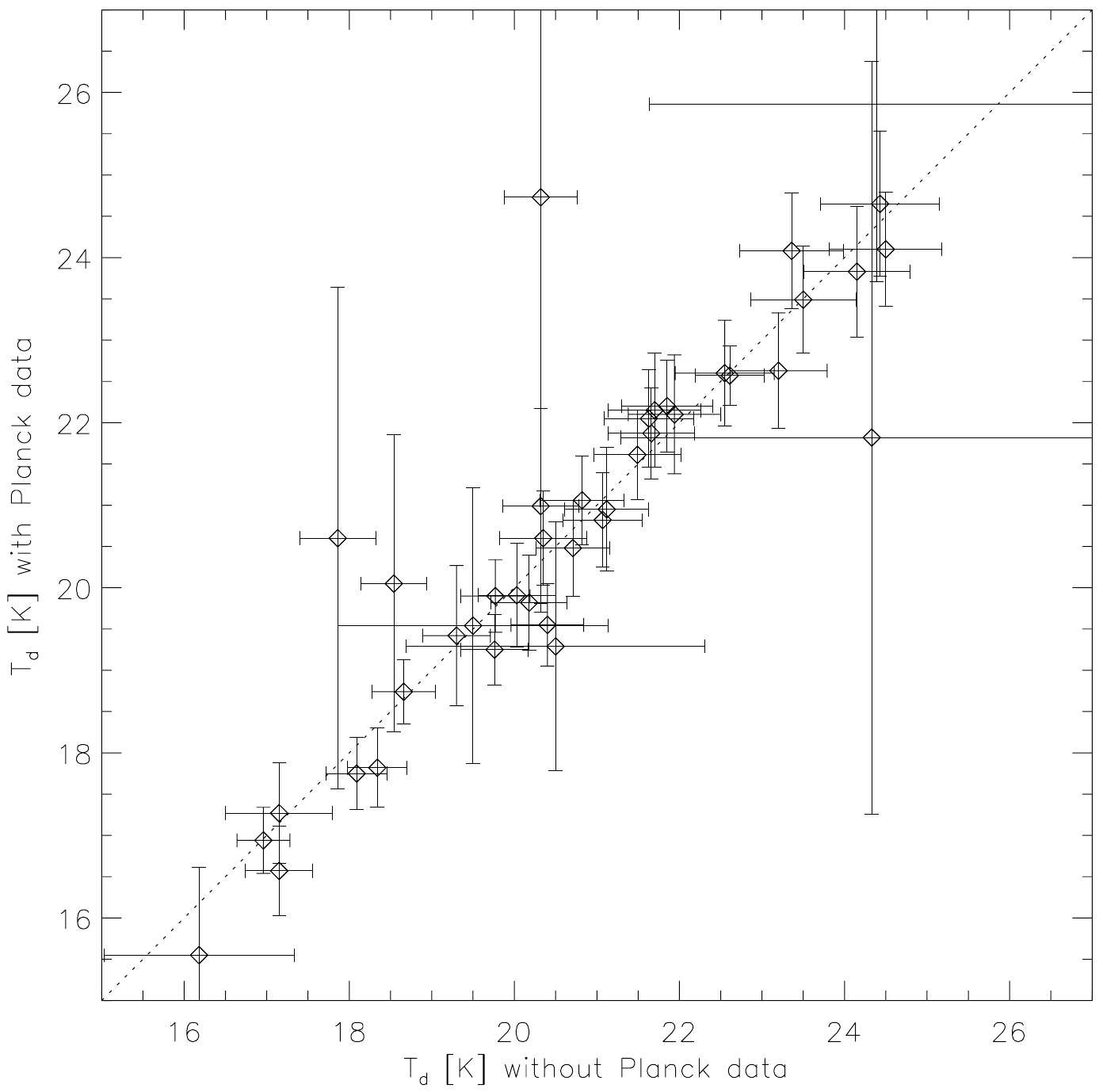}
\qquad
\includegraphics[width=0.35\textwidth,height=0.35\textwidth]{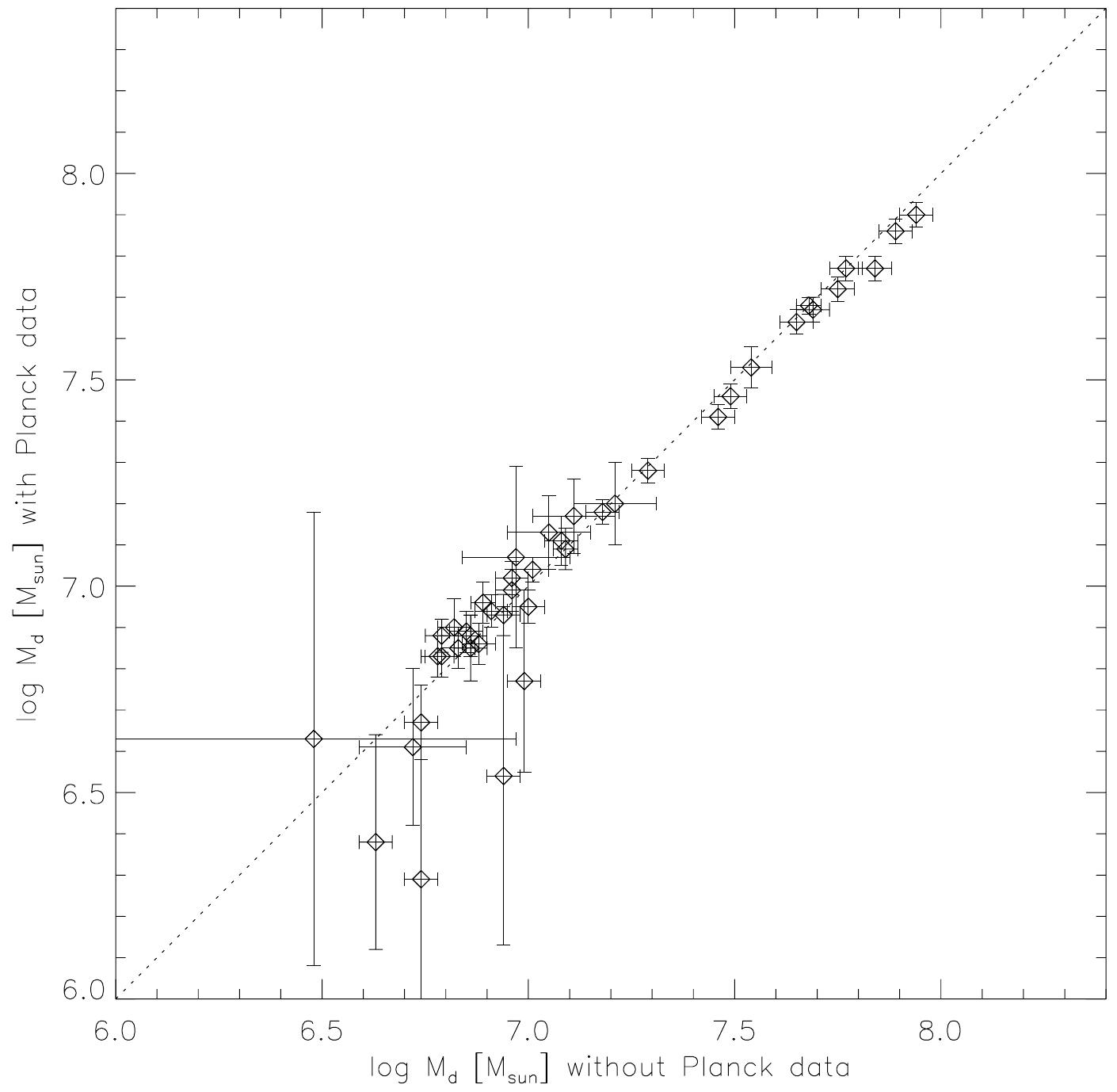}
\qquad
\caption{Comparison between the dust temperature and dust mass of 42 Virgo Cluster galaxies detected by Planck at 545 GHz, as derived from modified blackbody fits to the Herschel data alone (horizontal axis) and the combined Herschel and Planck 857 and 545 data (vertical axis). The dotted line corresponds to a one-to-one relation.}
\label{ComparePlanck.pdf}
\end{figure*}

Fitting modified blackbody fits to the PACS, SPIRE and HFI APERFLUX data, we find a mean temperature $\langle T_{\text{d}}\rangle = (21.3\pm3.2)$~K and dust mass $\langle\log M_{\text{d}}\rangle = 7.10\pm0.17$. These values are consistent with previous studies of the dust properties in late-type galaxies in the Virgo Cluster \citep{2012MNRAS.419.3505D, 2013arXiv1311.1774D, 2013MNRAS.428.1880A}, especially if we take into account that sometimes other assumptions have been made on the value of $\beta$ and the distances. More important here than the absolute values, however, is whether the inclusion of Planck data has a systematic effect on the derived dust characteristics. Figure~{\ref{ComparePlanck.pdf}} compares the dust mass and temperature for all galaxies in our sample, fitted with and without the Planck 857 and 545 GHz flux densities. The consistency between both fits is convincing: for every individual galaxy in our sample, the fitted modified blackbody parameters are fully consistent within the error bars. This implies that the addition or omission of the Planck 857 and 545 GHz flux densities does not imply a systematic bias, which is not unexpected given the good agreement between the SPIRE and Planck HFI flux densities (Section~\ref{FluxComparison.sec}).

Note that we have, at this stage, limited ourselves to the 857 and 545 GHz data, whereas Planck covers wavelengths up to the mm range. Our conclusion that the addition of 857 and 545 GHz data to the Herschel PACS and SPIRE data points does not affect the derived dust parameters does not necessarily imply that the same applies when longer wavelength Planck data are also taken into account. In particular, there has been quite some evidence for a submm/mm excess beyond a simple modified blackbody, in particular for dwarf galaxies \citep[e.g.,][]{2003A&A...407..159G, 2005A&A...434..867G, 2011A&A...536A..88G, 2009A&A...508..645G, 2011A&A...532A..56G, 2010A&A...519A..67I, 2011A&A...536A..17P, 2012ApJ...745...95D}. A study of the population of Planck sources detected at longer wavelengths, including their extended spectral energy distribution, the implied dust properties and a possible submm/mm excess, will be considered in a separate HeViCS study.

\subsection{Nature of the sources without counterparts}
\label{NoCounterpart.sec}

The PCCS contains two 857 GHz sources and five 545 GHz sources without an obvious nearby galaxy as counterpart. These Planck sources are particularly interesting as their submm flux density might originate from the common contribution of intermediate redshift galaxies or high-redshift ultra-luminous infrared galaxies. An example of such a compact 857 GHz Planck source is PLCKERC857 G270.59+58.52, linked to the strongly lensed hyperluminous starburst galaxy H-ATLAS J114637.9+001132 \citep{2012ApJ...753..134F}. 


\begin{figure*}
\centering
\includegraphics[width=\textwidth]{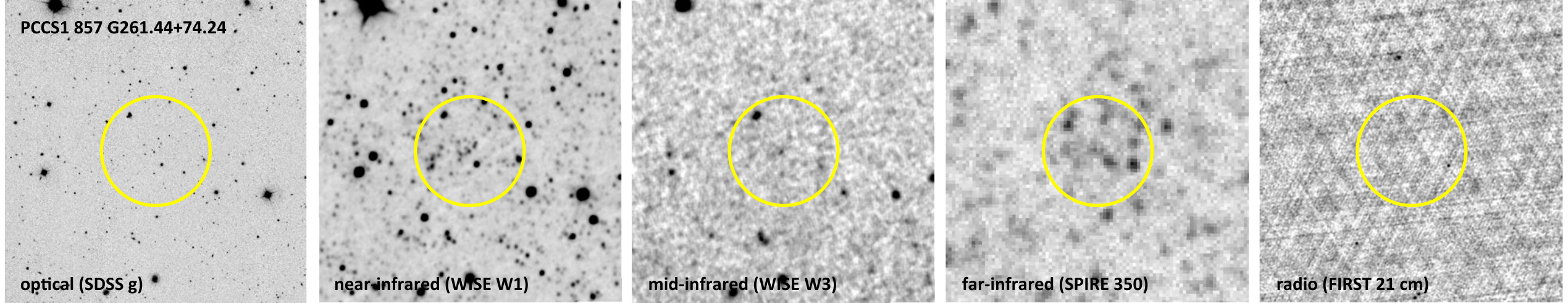}\\[1mm]%
\includegraphics[width=\textwidth]{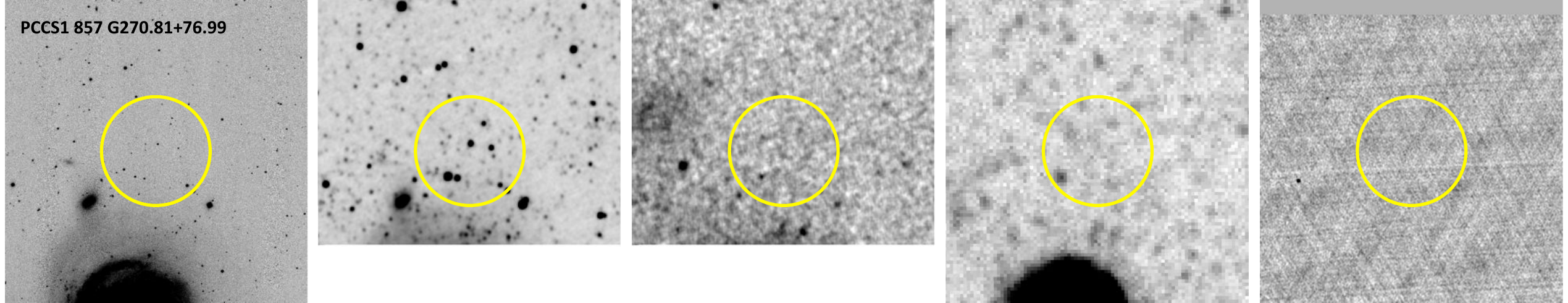}\\[1mm]%
\includegraphics[width=\textwidth]{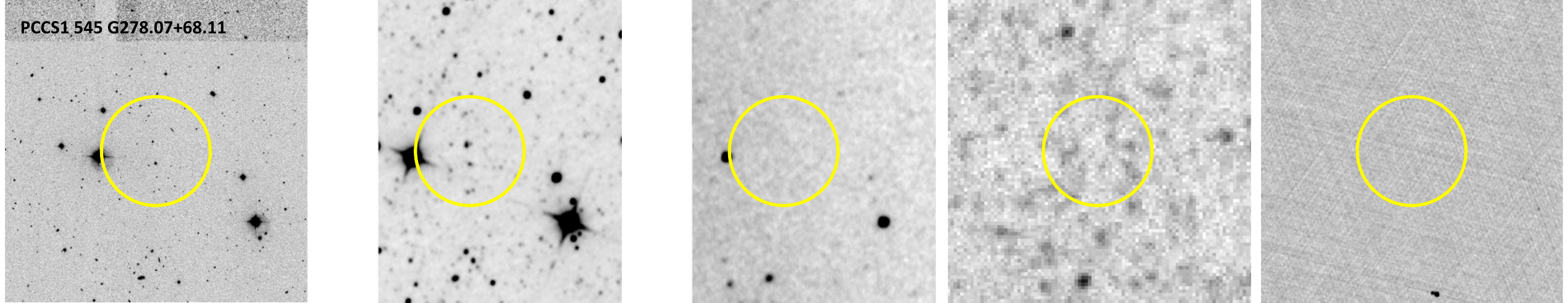}\\[1mm]%
\includegraphics[width=\textwidth]{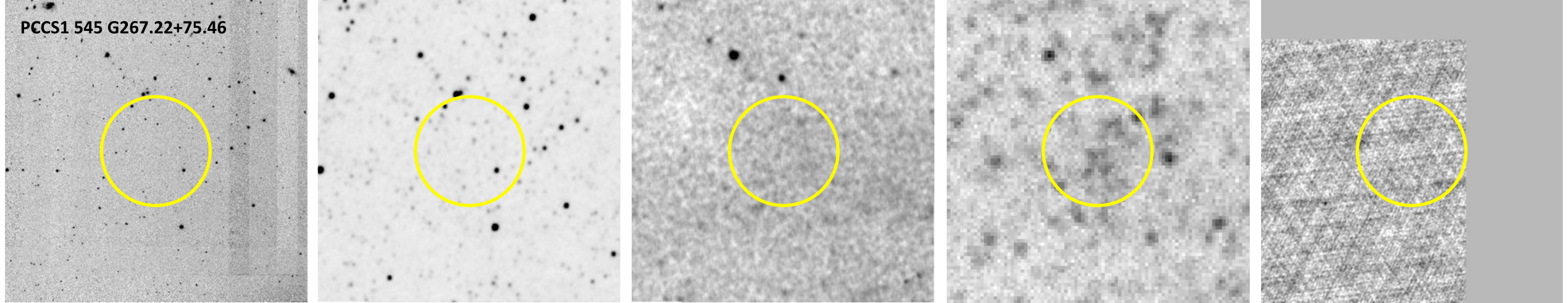}
\caption{Same as Figure~{\ref{DetectedSources.pdf}}, but now for four Planck sources without an obvious counterpart at optical wavelengths or in the Herschel maps.}
\label{UndetectedSources.pdf}
\end{figure*}

\begin{figure*}
\centering
\includegraphics[width=1.2\columnwidth]{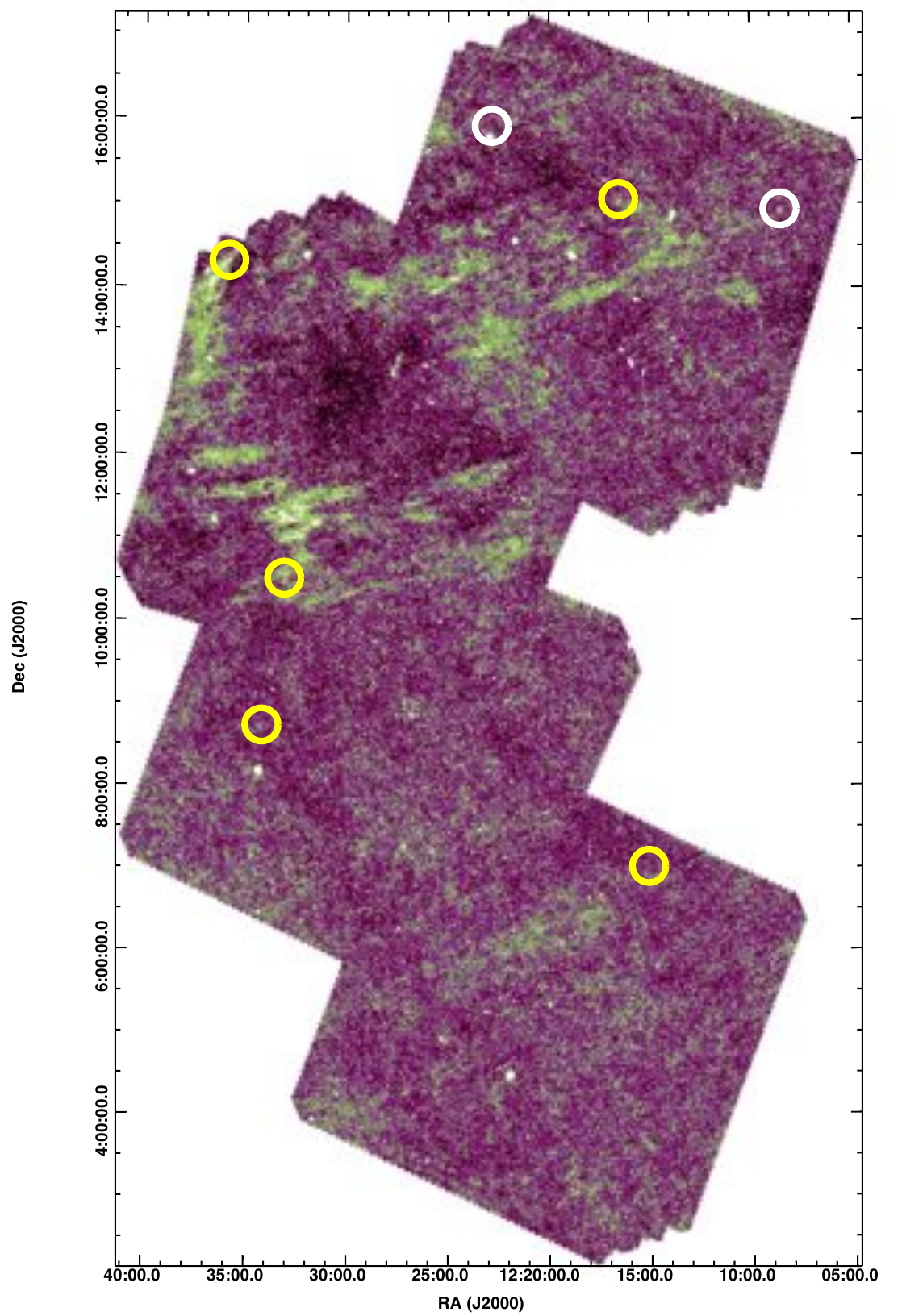}
\caption{Location of the PCCS sources without obvious SDSS or HeViCS counterpart on the HeViCS SPIRE 350 $\mu$m map. The white circles correspond to the 857 GHz sources, the yellow circles to 545 GHz sources. The SPIRE map has been smoothed to enhance the visibility of the Galactic cirrus features.}
\label{HeViCS-map.pdf}
\end{figure*}
 
Within the beam area around the position of the first unidentified 857 GHz source in the HeViCS fields, PCCS1 857 G261.44+74.24, NED lists 55 SDSS galaxies and one NVSS radio source. All of these galaxies are very dim at optical wavelengths: one galaxy has a $g$-band magnitude of 15.8, the others all are fainter than $m_g = 17.7$~mag. The top panel of Figure~{\ref{UndetectedSources.pdf}} compares the HeViCS SPIRE 350~$\mu$m image of the region around PCCS1 857 G261.44+74.24 with an optical SDSS image, near- and mid-infrared images from WISE and a radio continuum image from FIRST. The SPIRE image contains a number of discrete sources, all of them with flux densities below 75~mJy. Only one of the sources, with $S_{350}\sim40$~mJy, has an obvious galaxy as counterpart in the SDSS map, the $m_g=17.7$~mag galaxy SDSS J120835+145750 at redshift $z=0.083$. This Planck source is located at the edge of a prominent cirrus feature that extends from the centre of the cluster in the northwestern direction (Figure~{\ref{HeViCS-map.pdf}}), and it is marked as an extended source in the PCCS, i.e.\ it has the EXTENDED flag equal to one. Putting all this information together, it seems logical that the Planck detection corresponds to a Galactic cirrus feature (with potentially a non-negligible contribution from background galaxies).

The second unidentified 857 GHz PCCS source, PCCS1 857 G270.81+76.99 (Figure~{\ref{UndetectedSources.pdf}}, second row), is located in a region less affected by Galactic cirrus and is not marked as an extended source in the PCCS. NED lists 37 sources with the beam centered around the position of the source, most of them SDSS galaxies fainter than $m_g = 19$~mag. This might hence be a more promising candidate for an intermediate or high-redshift counterpart. However, the source is located very close ($<3$~arcmin) to M100, with a flux density of almost 26~Jy the brightest source in the Virgo Cluster at 350~$\mu$m \citep{2013MNRAS.428.1880A}. As there are no SPIRE 350~$\mu$m sources with flux densities above 50 mJy within the Planck beam at this position (second row on Figure {\ref{UndetectedSources.pdf}}), we presume that this source is a spurious detection.

The five 545 GHz PCCS sources without obvious optical counterpart are different from the two 857 GHz sources without optical counterpart. Looking at the position of these sources on the HeViCS map (Figure~{\ref{HeViCS-map.pdf}}), it seems obvious for at least two sources (PCCS1 545 G287.49+72.83 and G286.36+76.69) that they are associated with Galactic cirrus emission, as they are located on top of the notorious Galactic cirrus ring that surrounds the central regions of the Virgo Cluster \citep[e.g.,][]{2005ApJ...631L..41M,  2010ApJ...720..569R, 2010MNRAS.403L..26C}. Also for the other sources, we expect them to be dominated by cirrus emission. In all cases, the corresponding fields in the HeViCS SPIRE maps show a collection of faint sources, none of them bright enough to contribute a substantial part of the submm flux density, and most of them not associated to obvious sources in the corresponding SDSS optical maps (bottom two rows of Figure~{\ref{UndetectedSources.pdf}}).

Altogether, the number of PCCS sources without optical counterpart is very modest, especially at 857 GHz (2 out of 84). This is particularly striking when we compare these numbers to the results from the equatorial fields of the H-ATLAS survey by \citet{2013A&A...549A..31H}. Of the 28 Planck sources they found at 857 GHz, only 11 correspond to bright low-redshift galaxies (one of them a pair of nearby galaxies) and as many as 17 have no nearby galaxy as counterpart. Of these 17 unidentified sources, one source is resolved into a condensation of high-redshift point sources clustered around a strongly lensed galaxy, while the remaining 16 are probably related to Galactic cirrus features. A similar situation, but with fewer detected sources, is seen at 545 GHz. 

Even though it is logical that we detect many more nearby galaxies compared to \citet{2013A&A...549A..31H} (as we are looking at a nearby cluster), the number of sources without optical counterpart remains modest. Based on the number of sources and the area of the H-ATLAS equatorial fields (134.55 deg$^2$), we would expect 10 to 11 sources at 857 GHz without obvious optical counterpart. The obvious reason that explains this difference is the high Galactic latitude ($b=67-76$~deg) of the HeViCS fields compared to the equatorial H-ATLAS fields. In particular, all but one of the Planck detections by \citet{2013A&A...549A..31H} are located in the GAMA-09 field, located at a Galactic latitude around 30 deg and known to be strongly contaminated by cirrus \citep{2011MNRAS.412.1151B}. In spite of the clear signatures of cirrus in the HeViCS fields (see Figure~{\ref{HeViCS-map.pdf}}), this difference in Galactic latitude is probably responsible for the much lower number of Planck cirrus features. An additional factor could have been the difference between the Planck catalogues used for both studies: we use the PCCS, whereas \citet{2013A&A...549A..31H} use the ERCSC. Compared to the ERCSC, the PCCS uses a different extraction method that is better at excluding slightly extended objects.

Finally, we did not find any Planck source in the HeViCS fields that corresponds to a proto-cluster of intermediate or high-redshift dusty galaxies or strongly lensed submm sources. Based on the extrapolation of the \citet{2013A&A...549A..31H} results, at most one such source would be expected. One factor to take into account is again the source extraction method used by PCCS, which is better at excluding slightly extended objects (not only cirrus, but also high-redshift clumps). 

We can also attempt to make an estimate based on theoretical studies whether it is expected that no high-redshift Planck sources are detected in the HeViCS fields. \citet{2005MNRAS.358..869N} made predictions on the surface density of proto-clusters of dusty galaxies detectable by Planck. For the evolution indicated by simulations, they expect a surface density at the Planck detection limit below $10^{-2}$~deg$^{-2}$. This corresponds to at most one detected proto-cluster in the HeViCS fields. These predictions are, however, endowed with very large uncertainties both because the predicted counts of proto-clusters are very steep (and therefore are critically dependent on the detection limit which is hard to determine because the protoclusters are expected to be resolved by Planck), and because there is a considerable uncertainty on the evolution of key quantities such as the amplitude of the three-point correlation function. It is expected that, for luminous matter, this amplitude behaves as $b^{-1}$ or $b^{-2}$, where $b$ is the bias parameter \citep{1993ApJ...413..447F, 2001ApJ...548..114S}, and in this case the expected surface density decreases by more than an order of magnitude compared to the case without three-point correlation amplitude redshift evolution. Moreover, due to the relatively small field size and the effect of cosmic variance, it is difficult to make a firm estimate of the expected number of sources. On the other hand, the expected number counts of the sources as a function of flux density are quite steep \citep{2005MNRAS.358..869N}, so it is likely that substantially more submm bright proto-clusters can be found at fainter fluxes.

\citet{2007MNRAS.377.1557N} and \citet{2012ApJ...755...46L} present predictions of the surface density of lensed spheroids. Constrained by the number of detected and candidate strongly lensed submm galaxies in the H-ATLAS Science Demonstration Field \citep{2010Sci...330..800N, 2012ApJ...749...65G}, the models of \citet{2012ApJ...755...46L} predict a surface density around $8\times10^{-4}$ deg$^{-2}$ at a 500 $\mu$m flux density limit of 1~Jy. Again, this estimation is very uncertain and could in principle be considered as a lower limit, as the number of lensed spheroids decreases very quickly with flux density and the well-known clustering of submm sources \citep{2010A&A...518L..11M, 2011MNRAS.412L..93S, 2012MNRAS.426.3455V} is not taken into account in these estimates. In any case, these numbers are so low that we do not expect any high-$z$ lensed systems in the 84 deg$^2$ HeViCS fields above the Planck detection limit.

\subsection{Completeness and positional accuracy of the PCCS}

According to \citet{2013arXiv1303.5088P}, the PCCS catalogue should be 90\% complete at 857~GHz down to a flux density level of 680~mJy. To verify this, we checked the SPIRE 350~$\mu$m flux densities of all Virgo Cluster galaxies in the catalogue of \citet{2013MNRAS.428.1880A}. This catalogue contains 72 galaxies with a $S_{350}>680$~mJy (taking into account the correction factors for the old beam size, extended flux density correction and updated SPIRE calibration). 69 of these sources are detected by Planck in the PCCS (as 67 sources, as two pairs of galaxies are blended into single Planck sources). The three non-detected sources (IC\,3061, NGC\,4351, and NGC\,4435) all have SPIRE 350 $\mu$m flux densities below 720 mJy. Below $S_{350}\sim650$~mJy the detection rate drops quickly, although there are still occasional PCCS detections with even lower SPIRE flux densities. The most extreme case is the peculiar Sa galaxy NGC\,4506 with $S_{350}=178$~mJy, which stands out in Figure~{\ref{FluxComparison857.pdf}} as the isolated point at the extreme left of the panels. Flux boosting and confusion with Galactic cirrus emission are probably responsible for the fact that this galaxy made it to the PCCS: it is located on top of the previously mentioned prominent cirrus ring in the centre of the Virgo Cluster.

The PCCS 545 GHz catalogue is reported to have a 90\% completeness limit of 570 mJy \citep{2013arXiv1303.5088P}. Using the conversion factor 0.83 as derived in Section 3, this corresponds to a 500 $\mu$m flux density level of 687 mJy. There are 37 galaxies in the \citet{2013MNRAS.428.1880A} catalogue with a SPIRE 500 $\mu$m flux density above 687 mJy (again taking into account the previously mentioned correction factors). All but one of these sources, the edge-on spiral NGC\,4316 with $S_{500}=694$ mJy, are detected by Planck at 545 GHz. The faintest source detected has $S_{500} = 502$~mJy. 

Overall, our results are in good agreement with the PCCS completeness estimates: in the HeViCS fields, the PCCS recovered $\sim$95\% of the sources with $S_{350}>680$~mJy and $\sim$95\% of the sources with $S_{550}>570$~mJy. This is the first empirical confirmation of the estimated completeness of the PCCS. For the HFI bands, the PCCS completeness estimates were based on simulations in which unresolved point sources of different intrinsic flux density levels were injected in the real maps and convolved with the effective beam. The fact that we confirm these numbers is a strong support for the internal PCCS validation procedure.

\begin{figure}
\includegraphics[width=0.46\textwidth]{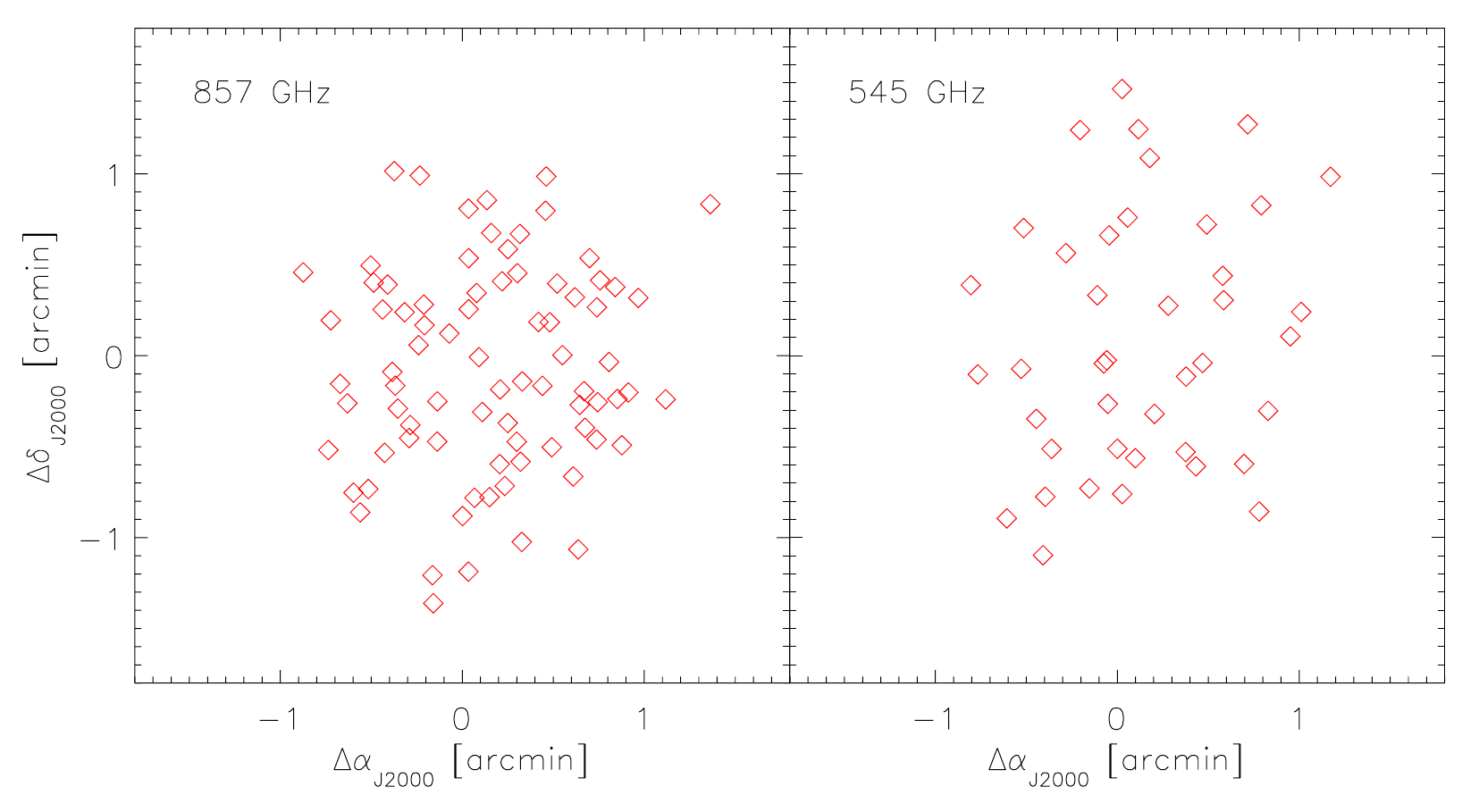}
\caption{Positional accuracy of the PCCS at 857 GHz (left panel) and 545 GHz (right panel). The plots show the difference in right ascension and declination between the position of the PCCS source and the coordinates of the centre of the optical counterpart, for all sources with a single nearby galaxy as counterpart.}
\label{PositionalAccuracy.pdf}
\end{figure}

Finally, in order to test the positional accuracy of the PCCS, we checked the positions of the sources as listed in the PCCS catalogue (and listed in Tables 1 and 2) with the central positions of the galaxies obtained from NED. The  results are shown in Figure~{\ref{PositionalAccuracy.pdf}}. There is no systematic offset, and the general positional accuracy is $0.71\pm0.29$ arcmin for the 857 GHz band, and $0.79\pm0.36$ arcmin for the 545 GHz band. These numbers are some 10\% larger than, but still fully consistent with, the positional uncertainties quoted by \citep{2013arXiv1303.5088P} based on the PCCS sources from the KINGFISH, HRS, H-ATLAS and HeViCS BGS samples (0.65 at 857 GHz and 0.72 at 545 GHz).

\section{Conclusion and summary}

We have cross-correlated the Planck Catalogue of Compact Sources (PCCS) with the 84 deg$^2$ survey fields of the Herschel Virgo Cluster Survey (HeViCS), with the goals of identifying and characterizing the counterparts of the compact Planck sources at 857 and 545 GHz, and searching for possible high-redshift gravitational lenses or proto-clusters of star-forming galaxies. The results of this investigation are the following:
\begin{enumerate}
\item We find 84 compact Planck sources at 857 GHz in the HeViCS survey fields. 77 of these sources correspond to individual Virgo Cluster galaxies, 3 to nearby background galaxies, and 2 to galaxy pairs in the Virgo Cluster. Only 2 sources do not have a clearly identifiable counterpart at optical and FIR/submm wavelengths. 
\item At 545 GHz, where the dominant emission mechanism is still continuum emission by cold and large dust grains, we find 48 sources in the PCCS. 41 of these sources have a single Virgo Cluster galaxy as counterpart, two correspond to a Virgo Cluster galaxy pair, and 5 sources have no nearby galaxy as a clearly identifiable counterpart at optical or submm wavelengths. No background galaxies are detected, and all sources with a counterpart, including the two galaxy pairs, are also detected at 857 GHz. 
\item The vast majority of all Planck detected galaxies are late-type spiral galaxies, with the Sc class dominating the numbers (more than half of all galaxies detected at 545 GHz). The number of early-type spirals reduces sharply when moving from 857 to 545 GHz: of the 43 galaxies detected at 545 GHz, there is only one Sa galaxy, compared to 10 out of 82 at the 857 GHz. Early-type galaxies are virtually absent from the PCCS, with only three lenticular galaxies (NGC\,4370, 4429 and 4526) and one elliptical galaxy (M87) detected. In the latter galaxy, detected at both 857 and 545 GHz, the submm emission is dominated by synchrotron rather than dust emission.
\item The PCCS presents four different estimates for the flux density, reflecting the different methods used to detect sources and measure their flux density. We have compared these different flux density measures to the SPIRE 350 and 500 $\mu$m flux density measures from the HeViCS Virgo Cluster catalogue \citep{2013MNRAS.428.1880A}. The Planck pipeline source detection algorithm estimate, DETFLUX, which showed the smallest dispersion in the comparison of \citet{2013arXiv1303.5088P}, shows systematic deviations both at the highest and the lowest flux densities. We find the best correlation between the SPIRE flux densities and the Planck aperture photometry flux densities (APERFLUX), even at the highest flux density levels. At 857 GHz, a linear regression fit results in a reduced $\chi^2$ compatible with a relation without intrinsic scatter. 
\item Based on the number of detected Virgo Cluster galaxies, we have estimated the completeness and positional accuracy of the PCCS at 857 and 545 GHz. Our estimates are in agreement with the results from  \citet{2013arXiv1303.5088P}, who report 90\% completeness limits of 680 and 570 mJy at 857 and 545 GHz respectively. Our study is the first empirical confirmation of the claimed completeness of the PCCS at submm wavelengths, which was estimated based on simulations, and hence provides a strong support for the internal PCCS validation procedure.
\item We have found only seven PCCS sources in the HeViCS fields without a nearby galaxy as obvious counterpart (two sources at 857 GHz and five sources at 545 GHz). Based on their location in the cluster and the SDSS and HeViCS sources at their positions, we conclude that all of these are probably linked to Galactic cirrus features or spurious detections. The number of sources without counterpart is remarkably low compared to the study by \citet{2013A&A...549A..31H}, who found that more than half of the compact Planck sources in the H-ATLAS equatorial fields correspond to cirrus features. This might be due to a combination of the higher galactic latitude of the Virgo Cluster compared to the equatorial H-ATLAS fields and the different Planck catalogue used. Unlike \citet{2013A&A...549A..31H}, we find no Planck sources in the HeViCS fields associated to high-redshift proto-clusters of dusty galaxies or strongly lensed submm sources.
\end{enumerate}

\section*{Acknowledgements}

MB, JF and TH acknowledge the financial support of the Belgian Science Policy Office (BELSPO) through the PRODEX project "Herschel-PACS Guaranteed Time and Open Time Programs: Science Exploitation" (C90370). MC and GDZ acknowledge financial support from ASI/INAF Agreement I/072/09/0. MB, FA, IDL, GG, SV and JV  acknowledge the support from the Flemish Fund for Scientific Research (FWO-Vlaanderen).

SPIRE has been developed by a consortium of institutes led by Cardiff University (UK) and including Univ. Lethbridge (Canada); NAOC (China); CEA, LAM (France); IFSI, Univ. Padua (Italy); IAC (Spain); Stockholm Observatory (Sweden); Imperial College London, RAL, UCL-MSSL, UKATC, Univ. Sussex (UK); and Caltech, JPL, NHSC, Univ. Colorado (USA). This development has been supported by national funding agencies: CSA (Canada); NAOC (China); CEA, CNES, CNRS (France); ASI (Italy); MCINN (Spain); SNSB (Sweden); STFC (UK); and NASA (USA).

The development of Planck has been supported by: ESA; CNES and CNRS/INSU-IN2P3-INP (France); ASI, CNR, and INAF (Italy); NASA and DoE (USA); STFC and UKSA (UK); CSIC, MICINN and JA (Spain); Tekes, AoF and CSC (Finland); DLR and MPG (Germany); CSA (Canada); DTU Space (Denmark); SER/SSO (Switzerland); RCN (Norway); SFI (Ireland); FCT/MCTES (Portugal); and The development of Planck has been supported by: ESA; CNES and CNRS/INSU-IN2P3-INP (France); ASI, CNR, and INAF (Italy); NASA and DoE (USA); STFC and UKSA (UK); CSIC, MICINN and JA (Spain); Tekes, AoF and CSC (Finland); DLR and MPG (Germany); CSA (Canada); DTU Space (Denmark); SER/SSO (Switzerland); RCN (Norway); SFI (Ireland); FCT/MCTES (Portugal); and PRACE (EU).

Funding for the SDSS and SDSS-II has been provided by the Alfred P. Sloan Foundation, the Participating Institutions, the National Science Foundation, the U.S. Department of Energy, the National Aeronautics and Space Administration, the Japanese Monbukagakusho, the Max Planck Society, and the Higher Education Funding Council for England. The SDSS Web Site is www.sdss.org. The SDSS is managed by the Astrophysical Research Consortium for the Participating Institutions. The Participating Institutions are the American Museum of Natural History, Astrophysical Institute Potsdam, University of Basel, University of Cambridge, Case Western Reserve University, University of Chicago, Drexel University, Fermilab, the Institute for Advanced Study, the Japan Participation Group, Johns Hopkins University, the Joint Institute for Nuclear Astrophysics, the Kavli Institute for Particle Astrophysics and Cosmology, the Korean Scientist Group, the Chinese Academy of Sciences (LAMOST), Los Alamos National Laboratory, the Max-Planck-Institute for Astronomy (MPIA), the Max-Planck-Institute for Astrophysics (MPA), New Mexico State University, Ohio State University, University of Pittsburgh, University of Portsmouth, Princeton University, the United States Naval Observatory, and the University of Washington.

This research has made use of the NASA/IPAC Extragalactic Database (NED) which is operated by the Jet Propulsion Laboratory, California Institute of Technology, under contract with the National Aeronautics and Space Administration. This research has made use of NASA's Astrophysics Data System Bibliographic Services. This research has made use of SAOImage DS9, developed by Smithsonian Astrophysical Observatory.

\bibliographystyle{aa} 
\bibliography{planck}

\begin{thebibliography}{114}
\expandafter\ifx\csname natexlab\endcsname\relax\def\natexlab#1{#1}\fi

\bibitem[{{Ahn} {et~al.}(2012){Ahn}, {Alexandroff}, {Allende Prieto},
  {Anderson}, {Anderton}, {Andrews}, {Aubourg}, {Bailey}, {Balbinot}, {Barnes},
  \& et~al.}]{2012ApJS..203...21A}
{Ahn}, C.~P., {Alexandroff}, R., {Allende Prieto}, C., {et~al.} 2012, \apjs,
  203, 21

\bibitem[{{Auld} {et~al.}(2013){Auld}, {Bianchi}, {Smith}, {Davies}, {Bendo},
  {di Serego}, {Cortese}, {Baes}, {Bomans}, {Boquien}, {Boselli}, {Ciesla},
  {Clemens}, {Corbelli}, {De Looze}, {Fritz}, {Gavazzi}, {Pappalardo},
  {Grossi}, {Hunt}, {Madden}, {Magrini}, {Pohlen}, {Verstappen}, {Vlahakis},
  {Xilouris}, \& {Zibetti}}]{2013MNRAS.428.1880A}
{Auld}, R., {Bianchi}, S., {Smith}, M.~W.~L., {et~al.} 2013, \mnras, 428, 1880

\bibitem[{{Baes} {et~al.}(2010){Baes}, {Clemens}, {Xilouris}, {Fritz},
  {Cotton}, {Davies}, {Bendo}, {Bianchi}, {Cortese}, {de Looze}, {Pohlen},
  {Verstappen}, {B{\"o}hringer}, {Bomans}, {Boselli}, {Corbelli}, {Dariush},
  {di Serego Alighieri}, {Fadda}, {Garcia-Appadoo}, {Gavazzi}, {Giovanardi},
  {Grossi}, {Hughes}, {Hunt}, {Jones}, {Madden}, {Pierini}, {Sabatini},
  {Smith}, {Vlahakis}, \& {Zibetti}}]{2010A&A...518L..53B}
{Baes}, M., {Clemens}, M., {Xilouris}, E.~M., {et~al.} 2010, \aap, 518, L53

\bibitem[{{Becker} {et~al.}(1995){Becker}, {White}, \&
  {Helfand}}]{1995ApJ...450..559B}
{Becker}, R.~H., {White}, R.~L., \& {Helfand}, D.~J. 1995, \apj, 450, 559

\bibitem[{{Bendo} {et~al.}(2012){Bendo}, {Boselli}, {Dariush}, {Pohlen},
  {Roussel}, {Sauvage}, {Smith}, {Wilson}, {Baes}, {Cooray}, {Clements},
  {Cortese}, {Foyle}, {Galametz}, {Gomez}, {Lebouteiller}, {Lu}, {Madden},
  {Mentuch}, {O'Halloran}, {Page}, {Remy}, {Schulz}, \&
  {Spinoglio}}]{2012MNRAS.419.1833B}
{Bendo}, G.~J., {Boselli}, A., {Dariush}, A., {et~al.} 2012, \mnras, 419, 1833

\bibitem[{{Bendo} {et~al.}(2013){Bendo}, {Griffin}, {Bock}, {Conversi},
  {Dowell}, {Lim}, {Lu}, {North}, {Papageorgiou}, {Pearson}, {Pohlen},
  {Polehampton}, {Schulz}, {Shupe}, {Sibthorpe}, {Spencer}, {Swinyard},
  {Valtchanov}, \& {Xu}}]{2013MNRAS.433.3062B}
{Bendo}, G.~J., {Griffin}, M.~J., {Bock}, J.~J., {et~al.} 2013, \mnras, 433,
  3062

\bibitem[{{Bendo} {et~al.}(2010){Bendo}, {Wilson}, {Pohlen}, {Sauvage}, {Auld},
  {Baes}, {Barlow}, {Bock}, {Boselli}, {Bradford}, {Buat}, {Castro-Rodriguez},
  {Chanial}, {Charlot}, {Ciesla}, {Clements}, {Cooray}, {Cormier}, {Cortese},
  {Davies}, {Dwek}, {Eales}, {Elbaz}, {Galametz}, {Galliano}, {Gear}, {Glenn},
  {Gomez}, {Griffin}, {Hony}, {Isaak}, {Levenson}, {Lu}, {Madden},
  {O'Halloran}, {Okumura}, {Oliver}, {Page}, {Panuzzo}, {Papageorgiou},
  {Parkin}, {Perez-Fournon}, {Rangwala}, {Rigby}, {Roussel}, {Rykala},
  {Sacchi}, {Schulz}, {Schirm}, {Smith}, {Spinoglio}, {Stevens}, {Sundar},
  {Symeonidis}, {Trichas}, {Vaccari}, {Vigroux}, {Wozniak}, {Wright}, \&
  {Zeilinger}}]{2010A&A...518L..65B}
{Bendo}, G.~J., {Wilson}, C.~D., {Pohlen}, M., {et~al.} 2010, \aap, 518, L65

\bibitem[{{Bersanelli} {et~al.}(2010){Bersanelli}, {Mandolesi}, {Butler},
  {Mennella}, {Villa}, {Aja}, {Artal}, {Artina}, {Baccigalupi}, {Balasini},
  {Baldan}, {Banday}, {Bastia}, {Battaglia}, {Bernardino}, {Blackhurst},
  {Boschini}, {Burigana}, {Cafagna}, {Cappellini}, {Cavaliere}, {Colombo},
  {Crone}, {Cuttaia}, {D'Arcangelo}, {Danese}, {Davies}, {Davis}, {de Angelis},
  {de Gasperis}, {de La Fuente}, {de Rosa}, {de Zotti}, {Falvella}, {Ferrari},
  {Ferretti}, {Figini}, {Fogliani}, {Franceschet}, {Franceschi}, {Gaier},
  {Garavaglia}, {Gomez}, {Gorski}, {Gregorio}, {Guzzi}, {Herreros},
  {Hildebrandt}, {Hoyland}, {Hughes}, {Janssen}, {Jukkala}, {Kettle},
  {Kilpi{\"a}}, {Laaninen}, {Lapolla}, {Lawrence}, {Lawson}, {Leahy},
  {Leonardi}, {Leutenegger}, {Levin}, {Lilje}, {Lowe}, {Lubin}, {Maino},
  {Malaspina}, {Maris}, {Marti-Canales}, {Martinez-Gonzalez}, {Mediavilla},
  {Meinhold}, {Miccolis}, {Morgante}, {Natoli}, {Nesti}, {Pagan}, {Paine},
  {Partridge}, {Pascual}, {Pasian}, {Pearson}, {Pecora}, {Perrotta},
  {Platania}, {Pospieszalski}, {Poutanen}, {Prina}, {Rebolo}, {Roddis},
  {Rubi{\~n}o-Martin}, {Salmon}, {Sandri}, {Seiffert}, {Silvestri},
  {Simonetto}, {Sjoman}, {Smoot}, {Sozzi}, {Stringhetti}, {Taddei}, {Tauber},
  {Terenzi}, {Tomasi}, {Tuovinen}, {Valenziano}, {Varis}, {Vittorio}, {Wade},
  {Wilkinson}, {Winder}, {Zacchei}, \& {Zonca}}]{2010A&A...520A...4B}
{Bersanelli}, M., {Mandolesi}, N., {Butler}, R.~C., {et~al.} 2010, \aap, 520,
  A4

\bibitem[{{Bianchi}(2013)}]{2013A&A...552A..89B}
{Bianchi}, S. 2013, \aap, 552, A89

\bibitem[{{Binggeli} {et~al.}(1985){Binggeli}, {Sandage}, \&
  {Tammann}}]{1985AJ.....90.1681B}
{Binggeli}, B., {Sandage}, A., \& {Tammann}, G.~A. 1985, \aj, 90, 1681

\bibitem[{{B{\"o}hringer} {et~al.}(1994){B{\"o}hringer}, {Briel}, {Schwarz},
  {Voges}, {Hartner}, \& {Tr{\"u}mper}}]{1994Natur.368..828B}
{B{\"o}hringer}, H., {Briel}, U.~G., {Schwarz}, R.~A., {et~al.} 1994, \nat,
  368, 828

\bibitem[{{Boquien} {et~al.}(2011){Boquien}, {Calzetti}, {Combes}, {Henkel},
  {Israel}, {Kramer}, {Rela{\~n}o}, {Verley}, {van der Werf}, {Xilouris}, \&
  {HERM33ES Team}}]{2011AJ....142..111B}
{Boquien}, M., {Calzetti}, D., {Combes}, F., {et~al.} 2011, \aj, 142, 111

\bibitem[{{Boselli} {et~al.}(2011){Boselli}, {Boissier}, {Heinis}, {Cortese},
  {Ilbert}, {Hughes}, {Cucciati}, {Davies}, {Ferrarese}, {Giovanelli},
  {Haynes}, {Baes}, {Balkowski}, {Brosch}, {Chapman}, {Charmandaris},
  {Clemens}, {Dariush}, {de Looze}, {di Serego Alighieri}, {Duc}, {Durrell},
  {Emsellem}, {Erben}, {Fritz}, {Garcia-Appadoo}, {Gavazzi}, {Grossi},
  {Jord{\'a}n}, {Hess}, {Huertas-Company}, {Hunt}, {Kent}, {Lambas}, {Lan{\c
  c}on}, {MacArthur}, {Madden}, {Magrini}, {Mei}, {Momjian}, {Olowin},
  {Papastergis}, {Smith}, {Solanes}, {Spector}, {Spekkens}, {Taylor},
  {Valotto}, {van Driel}, {Verstappen}, {Vlahakis}, {Vollmer}, \&
  {Xilouris}}]{2011A&A...528A.107B}
{Boselli}, A., {Boissier}, S., {Heinis}, S., {et~al.} 2011, \aap, 528, A107

\bibitem[{{Boselli} {et~al.}(2010{\natexlab{a}}){Boselli}, {Ciesla}, {Buat},
  {Cortese}, {Auld}, {Baes}, {Bendo}, {Bianchi}, {Bock}, {Bomans}, {Bradford},
  {Castro-Rodriguez}, {Chanial}, {Charlot}, {Clemens}, {Clements}, {Corbelli},
  {Cooray}, {Cormier}, {Dariush}, {Davies}, {de Looze}, {di Serego Alighieri},
  {Dwek}, {Eales}, {Elbaz}, {Fadda}, {Fritz}, {Galametz}, {Galliano},
  {Garcia-Appadoo}, {Gavazzi}, {Gear}, {Giovanardi}, {Glenn}, {Gomez},
  {Griffin}, {Grossi}, {Hony}, {Hughes}, {Hunt}, {Isaak}, {Jones}, {Levenson},
  {Lu}, {Madden}, {O'Halloran}, {Okumura}, {Oliver}, {Page}, {Panuzzo},
  {Papageorgiou}, {Parkin}, {Perez-Fournon}, {Pierini}, {Pohlen}, {Rangwala},
  {Rigby}, {Roussel}, {Rykala}, {Sabatini}, {Sacchi}, {Sauvage}, {Schulz},
  {Schirm}, {Smith}, {Spinoglio}, {Stevens}, {Sundar}, {Symeonidis}, {Trichas},
  {Vaccari}, {Verstappen}, {Vigroux}, {Vlahakis}, {Wilson}, {Wozniak},
  {Wright}, {Xilouris}, {Zeilinger}, \& {Zibetti}}]{2010A&A...518L..61B}
{Boselli}, A., {Ciesla}, L., {Buat}, V., {et~al.} 2010{\natexlab{a}}, \aap,
  518, L61

\bibitem[{{Boselli} {et~al.}(2012){Boselli}, {Ciesla}, {Cortese}, {Buat},
  {Boquien}, {Bendo}, {Boissier}, {Eales}, {Gavazzi}, {Hughes}, {Pohlen},
  {Smith}, {Baes}, {Bianchi}, {Clements}, {Cooray}, {Davies}, {Gear}, {Madden},
  {Magrini}, {Panuzzo}, {Remy}, {Spinoglio}, \&
  {Zibetti}}]{2012A&A...540A..54B}
{Boselli}, A., {Ciesla}, L., {Cortese}, L., {et~al.} 2012, \aap, 540, A54

\bibitem[{{Boselli} {et~al.}(2010{\natexlab{b}}){Boselli}, {Eales}, {Cortese},
  {Bendo}, {Chanial}, {Buat}, {Davies}, {Auld}, {Rigby}, {Baes}, {Barlow},
  {Bock}, {Bradford}, {Castro-Rodriguez}, {Charlot}, {Clements}, {Cormier},
  {Dwek}, {Elbaz}, {Galametz}, {Galliano}, {Gear}, {Glenn}, {Gomez}, {Griffin},
  {Hony}, {Isaak}, {Levenson}, {Lu}, {Madden}, {O'Halloran}, {Okamura},
  {Oliver}, {Page}, {Panuzzo}, {Papageorgiou}, {Parkin}, {Perez-Fournon},
  {Pohlen}, {Rangwala}, {Roussel}, {Rykala}, {Sacchi}, {Sauvage}, {Schulz},
  {Schirm}, {Smith}, {Spinoglio}, {Stevens}, {Symeonidis}, {Vaccari},
  {Vigroux}, {Wilson}, {Wozniak}, {Wright}, \&
  {Zeilinger}}]{2010PASP..122..261B}
{Boselli}, A., {Eales}, S., {Cortese}, L., {et~al.} 2010{\natexlab{b}}, \pasp,
  122, 261

\bibitem[{{Bracco} {et~al.}(2011){Bracco}, {Cooray}, {Veneziani}, {Amblard},
  {Serra}, {Wardlow}, {Thompson}, {White}, {Auld}, {Baes}, {Bertoldi},
  {Buttiglione}, {Cava}, {Clements}, {Dariush}, {de Zotti}, {Dunne}, {Dye},
  {Eales}, {Fritz}, {Gomez}, {Hopwood}, {Ibar}, {Ivison}, {Jarvis}, {Lagache},
  {Lee}, {Leeuw}, {Maddox}, {Micha{\l}owski}, {Pearson}, {Pohlen}, {Rigby},
  {Rodighiero}, {Smith}, {Temi}, {Vaccari}, \& {van der
  Werf}}]{2011MNRAS.412.1151B}
{Bracco}, A., {Cooray}, A., {Veneziani}, M., {et~al.} 2011, \mnras, 412, 1151

\bibitem[{{Buson} {et~al.}(2009){Buson}, {Bressan}, {Panuzzo}, {Rampazzo},
  {Vald{\'e}s}, {Clemens}, {Marino}, {Chavez}, {Granato}, \&
  {Silva}}]{2009ApJ...705..356B}
{Buson}, L., {Bressan}, A., {Panuzzo}, P., {et~al.} 2009, \apj, 705, 356

\bibitem[{{Ciesla} {et~al.}(2012){Ciesla}, {Boselli}, {Smith}, {Bendo},
  {Cortese}, {Eales}, {Bianchi}, {Boquien}, {Buat}, {Davies}, {Pohlen},
  {Zibetti}, {Baes}, {Cooray}, {de Looze}, {di Serego Alighieri}, {Galametz},
  {Gomez}, {Lebouteiller}, {Madden}, {Pappalardo}, {Remy}, {Spinoglio},
  {Vaccari}, {Auld}, \& {Clements}}]{2012A&A...543A.161C}
{Ciesla}, L., {Boselli}, A., {Smith}, M.~W.~L., {et~al.} 2012, \aap, 543, A161

\bibitem[{{Clemens} {et~al.}(2013){Clemens}, {Negrello}, {De Zotti},
  {Gonzalez-Nuevo}, {Bonavera}, {Cosco}, {Guarese}, {Boaretto}, {Salucci},
  {Baccigalupi}, {Clements}, {Danese}, {Lapi}, {Mandolesi}, {Partridge},
  {Perrotta}, {Serjeant}, {Scott}, \& {Toffolatti}}]{2013MNRAS.433..695C}
{Clemens}, M.~S., {Negrello}, M., {De Zotti}, G., {et~al.} 2013, \mnras, 433,
  695

\bibitem[{{Clements} {et~al.}(2010){Clements}, {Rigby}, {Maddox}, {Dunne},
  {Mortier}, {Pearson}, {Amblard}, {Auld}, {Baes}, {Bonfield}, {Burgarella},
  {Buttiglione}, {Cava}, {Cooray}, {Dariush}, {de Zotti}, {Dye}, {Eales},
  {Frayer}, {Fritz}, {Gardner}, {Gonzalez-Nuevo}, {Herranz}, {Ibar}, {Ivison},
  {Jarvis}, {Lagache}, {Leeuw}, {Lopez-Caniego}, {Negrello}, {Pascale},
  {Pohlen}, {Rodighiero}, {Samui}, {Serjeant}, {Sibthorpe}, {Scott}, {Smith},
  {Temi}, {Thompson}, {Valtchanov}, {van der Werf}, \&
  {Verma}}]{2010A&A...518L...8C}
{Clements}, D.~L., {Rigby}, E., {Maddox}, S., {et~al.} 2010, \aap, 518, L8

\bibitem[{{Compi{\`e}gne} {et~al.}(2011){Compi{\`e}gne}, {Verstraete}, {Jones},
  {Bernard}, {Boulanger}, {Flagey}, {Le Bourlot}, {Paradis}, \&
  {Ysard}}]{2011A&A...525A.103C}
{Compi{\`e}gne}, M., {Verstraete}, L., {Jones}, A., {et~al.} 2011, \aap, 525,
  A103

\bibitem[{{Corbelli} {et~al.}(2012){Corbelli}, {Bianchi}, {Cortese},
  {Giovanardi}, {Magrini}, {Pappalardo}, {Boselli}, {Bendo}, {Davies},
  {Grossi}, {Madden}, {Smith}, {Vlahakis}, {Auld}, {Baes}, {De Looze}, {Fritz},
  {Pohlen}, \& {Verstappen}}]{2012A&A...542A..32C}
{Corbelli}, E., {Bianchi}, S., {Cortese}, L., {et~al.} 2012, \aap, 542, A32

\bibitem[{{Cortese} {et~al.}(2010{\natexlab{a}}){Cortese}, {Bendo}, {Isaak},
  {Davies}, \& {Kent}}]{2010MNRAS.403L..26C}
{Cortese}, L., {Bendo}, G.~J., {Isaak}, K.~G., {Davies}, J.~I., \& {Kent},
  B.~R. 2010{\natexlab{a}}, \mnras, 403, L26

\bibitem[{{Cortese} {et~al.}(2010{\natexlab{b}}){Cortese}, {Davies}, {Pohlen},
  {Baes}, {Bendo}, {Bianchi}, {Boselli}, {De Looze}, {Fritz}, {Verstappen},
  {Bomans}, {Clemens}, {Corbelli}, {Dariush}, {di Serego Alighieri}, {Fadda},
  {Garcia-Appadoo}, {Gavazzi}, {Giovanardi}, {Grossi}, {Hughes}, {Hunt},
  {Jones}, {Madden}, {Pierini}, {Sabatini}, {Smith}, {Vlahakis}, {Xilouris}, \&
  {Zibetti}}]{2010A&A...518L..49C}
{Cortese}, L., {Davies}, J.~I., {Pohlen}, M., {et~al.} 2010{\natexlab{b}},
  \aap, 518, L49

\bibitem[{{Cotton} {et~al.}(2009){Cotton}, {Mason}, {Dicker}, {Korngut},
  {Devlin}, {Aquirre}, {Benford}, {Moseley}, {Staguhn}, {Irwin}, \&
  {Ade}}]{2009ApJ...701.1872C}
{Cotton}, W.~D., {Mason}, B.~S., {Dicker}, S.~R., {et~al.} 2009, \apj, 701,
  1872

\bibitem[{{Dale} {et~al.}(2012){Dale}, {Aniano}, {Engelbracht}, {Hinz},
  {Krause}, {Montiel}, {Roussel}, {Appleton}, {Armus}, {Beir{\~a}o}, {Bolatto},
  {Brandl}, {Calzetti}, {Crocker}, {Croxall}, {Draine}, {Galametz}, {Gordon},
  {Groves}, {Hao}, {Helou}, {Hunt}, {Johnson}, {Kennicutt}, {Koda}, {Leroy},
  {Li}, {Meidt}, {Miller}, {Murphy}, {Rahman}, {Rix}, {Sandstrom}, {Sauvage},
  {Schinnerer}, {Skibba}, {Smith}, {Tabatabaei}, {Walter}, {Wilson}, {Wolfire},
  \& {Zibetti}}]{2012ApJ...745...95D}
{Dale}, D.~A., {Aniano}, G., {Engelbracht}, C.~W., {et~al.} 2012, \apj, 745, 95

\bibitem[{{Davies} {et~al.}(2004){Davies}, {Minchin}, {Sabatini}, {van Driel},
  {Baes}, {Boyce}, {de Blok}, {Disney}, {Evans}, {Kilborn}, {Lang}, {Linder},
  {Roberts}, \& {Smith}}]{2004MNRAS.349..922D}
{Davies}, J., {Minchin}, R., {Sabatini}, S., {et~al.} 2004, \mnras, 349, 922

\bibitem[{{Davies} {et~al.}(2010){Davies}, {Baes}, {Bendo}, {Bianchi},
  {Bomans}, {Boselli}, {Clemens}, {Corbelli}, {Cortese}, {Dariush}, {De Looze},
  {di Serego Alighieri}, {Fadda}, {Fritz}, {Garcia-Appadoo}, {Gavazzi},
  {Giovanardi}, {Grossi}, {Hughes}, {Hunt}, {Jones}, {Madden}, {Pierini},
  {Pohlen}, {Sabatini}, {Smith}, {Verstappen}, {Vlahakis}, {Xilouris}, \&
  {Zibetti}}]{2010A&A...518L..48D}
{Davies}, J.~I., {Baes}, M., {Bendo}, G.~J., {et~al.} 2010, \aap, 518, L48

\bibitem[{{Davies} {et~al.}(2013){Davies}, {Bianchi}, {Baes}, {Bendo},
  {Clemens}, {De Looze}, {di Serego Alighieri}, {Fritz}, {Fuller},
  {Pappalardo}, {Hughes}, {Madden}, {Smith}, {Verstappen}, \&
  {Vlahakis}}]{2013arXiv1311.1774D}
{Davies}, J.~I., {Bianchi}, S., {Baes}, M., {et~al.} 2013, \mnras, in press,
  arXiv:1311.1774

\bibitem[{{Davies} {et~al.}(2012){Davies}, {Bianchi}, {Cortese}, {Auld},
  {Baes}, {Bendo}, {Boselli}, {Ciesla}, {Clemens}, {Corbelli}, {De Looze},
  {Alighieri}, {Fritz}, {Gavazzi}, {Pappalardo}, {Grossi}, {Hunt}, {Madden},
  {Magrini}, {Pohlen}, {Smith}, {Verstappen}, \&
  {Vlahakis}}]{2012MNRAS.419.3505D}
{Davies}, J.~I., {Bianchi}, S., {Cortese}, L., {et~al.} 2012, \mnras, 419, 3505

\bibitem[{{de Graauw} {et~al.}(2010){de Graauw}, {Helmich}, {Phillips},
  {Stutzki}, {Caux}, {Whyborn}, {Dieleman}, {Roelfsema}, {Aarts}, {Assendorp},
  {Bachiller}, {Baechtold}, {Barcia}, {Beintema}, {Belitsky}, {Benz}, {Bieber},
  {Boogert}, {Borys}, {Bumble}, {Ca{\"i}s}, {Caris}, {Cerulli-Irelli},
  {Chattopadhyay}, {Cherednichenko}, {Ciechanowicz}, {Coeur-Joly}, {Comito},
  {Cros}, {de Jonge}, {de Lange}, {Delforges}, {Delorme}, {den Boggende},
  {Desbat}, {Diez-Gonz{\'a}lez}, {di Giorgio}, {Dubbeldam}, {Edwards},
  {Eggens}, {Erickson}, {Evers}, {Fich}, {Finn}, {Franke}, {Gaier}, {Gal},
  {Gao}, {Gallego}, {Gauffre}, {Gill}, {Glenz}, {Golstein}, {Goulooze},
  {Gunsing}, {G{\"u}sten}, {Hartogh}, {Hatch}, {Higgins}, {Honingh}, {Huisman},
  {Jackson}, {Jacobs}, {Jacobs}, {Jarchow}, {Javadi}, {Jellema}, {Justen},
  {Karpov}, {Kasemann}, {Kawamura}, {Keizer}, {Kester}, {Klapwijk}, {Klein},
  {Kollberg}, {Kooi}, {Kooiman}, {Kopf}, {Krause}, {Krieg}, {Kramer},
  {Kruizenga}, {Kuhn}, {Laauwen}, {Lai}, {Larsson}, {Leduc}, {Leinz}, {Lin},
  {Liseau}, {Liu}, {Loose}, {L{\'o}pez-Fernandez}, {Lord}, {Luinge}, {Marston},
  {Mart{\'{\i}}n-Pintado}, {Maestrini}, {Maiwald}, {McCoey}, {Mehdi}, {Megej},
  {Melchior}, {Meinsma}, {Merkel}, {Michalska}, {Monstein}, {Moratschke},
  {Morris}, {Muller}, {Murphy}, {Naber}, {Natale}, {Nowosielski}, {Nuzzolo},
  {Olberg}, {Olbrich}, {Orfei}, {Orleanski}, {Ossenkopf}, {Peacock}, {Pearson},
  {Peron}, {Phillip-May}, {Piazzo}, {Planesas}, {Rataj}, {Ravera}, {Risacher},
  {Salez}, {Samoska}, {Saraceno}, {Schieder}, {Schlecht}, {Schl{\"o}der},
  {Schm{\"u}lling}, {Schultz}, {Schuster}, {Siebertz}, {Smit}, {Szczerba},
  {Shipman}, {Steinmetz}, {Stern}, {Stokroos}, {Teipen}, {Teyssier}, {Tils},
  {Trappe}, {van Baaren}, {van Leeuwen}, {van de Stadt}, {Visser}, {Wildeman},
  {Wafelbakker}, {Ward}, {Wesselius}, {Wild}, {Wulff}, {Wunsch}, {Tielens},
  {Zaal}, {Zirath}, {Zmuidzinas}, \& {Zwart}}]{2010A&A...518L...6D}
{de Graauw}, T., {Helmich}, F.~P., {Phillips}, T.~G., {et~al.} 2010, \aap, 518,
  L6

\bibitem[{{De Looze} {et~al.}(2013){De Looze}, {Baes}, {Boselli}, {Cortese},
  {Fritz}, {Auld}, {Bendo}, {Bianchi}, {Boquien}, {Clemens}, {Ciesla},
  {Davies}, {di Serego Alighieri}, {Grossi}, {Jones}, {Madden}, {Pappalardo},
  {Pierini}, {Smith}, {Verstappen}, {Vlahakis}, \&
  {Zibetti}}]{2013MNRAS.436.1057D}
{De Looze}, I., {Baes}, M., {Boselli}, A., {et~al.} 2013, \mnras, 436, 1057

\bibitem[{{De Looze} {et~al.}(2010){De Looze}, {Baes}, {Zibetti}, {Fritz},
  {Cortese}, {Davies}, {Verstappen}, {Bendo}, {Bianchi}, {Clemens}, {Bomans},
  {Boselli}, {Corbelli}, {Dariush}, {di Serego Alighieri}, {Fadda},
  {Garcia-Appadoo}, {Gavazzi}, {Giovanardi}, {Grossi}, {Hughes}, {Hunt},
  {Jones}, {Madden}, {Pierini}, {Pohlen}, {Sabatini}, {Smith}, {Vlahakis}, \&
  {Xilouris}}]{2010A&A...518L..54D}
{De Looze}, I., {Baes}, M., {Zibetti}, S., {et~al.} 2010, \aap, 518, L54

\bibitem[{{de Vaucouleurs} {et~al.}(1991){de Vaucouleurs}, {de Vaucouleurs},
  {Corwin}, {Buta}, {Paturel}, \& {Fouqu{\'e}}}]{1991rc3..book.....D}
{de Vaucouleurs}, G., {de Vaucouleurs}, A., {Corwin}, Jr., H.~G., {et~al.}
  1991, {Third Reference Catalogue of Bright Galaxies} (Springer (New York))

\bibitem[{{di Serego Alighieri} {et~al.}(2013){di Serego Alighieri}, {Bianchi},
  {Pappalardo}, {Zibetti}, {Auld}, {Baes}, {Bendo}, {Corbelli}, {Davies},
  {Davis}, {De Looze}, {Fritz}, {Gavazzi}, {Giovanardi}, {Grossi}, {Hunt},
  {Magrini}, {Pierini}, \& {Xilouris}}]{2013A&A...552A...8D}
{di Serego Alighieri}, S., {Bianchi}, S., {Pappalardo}, C., {et~al.} 2013,
  \aap, 552, A8

\bibitem[{{Draine}(2003)}]{2003ARA&A..41..241D}
{Draine}, B.~T. 2003, \araa, 41, 241

\bibitem[{{Draine} \& {Li}(2007)}]{2007ApJ...657..810D}
{Draine}, B.~T. \& {Li}, A. 2007, \apj, 657, 810

\bibitem[{{Eales} {et~al.}(2010){Eales}, {Dunne}, {Clements}, {Cooray}, {de
  Zotti}, {Dye}, {Ivison}, {Jarvis}, {Lagache}, {Maddox}, {Negrello},
  {Serjeant}, {Thompson}, {van Kampen}, {Amblard}, {Andreani}, {Baes},
  {Beelen}, {Bendo}, {Benford}, {Bertoldi}, {Bock}, {Bonfield}, {Boselli},
  {Bridge}, {Buat}, {Burgarella}, {Carlberg}, {Cava}, {Chanial}, {Charlot},
  {Christopher}, {Coles}, {Cortese}, {Dariush}, {da Cunha}, {Dalton}, {Danese},
  {Dannerbauer}, {Driver}, {Dunlop}, {Fan}, {Farrah}, {Frayer}, {Frenk},
  {Geach}, {Gardner}, {Gomez}, {Gonz{\'a}lez-Nuevo}, {Gonz{\'a}lez-Solares},
  {Griffin}, {Hardcastle}, {Hatziminaoglou}, {Herranz}, {Hughes}, {Ibar},
  {Jeong}, {Lacey}, {Lapi}, {Lawrence}, {Lee}, {Leeuw}, {Liske},
  {L{\'o}pez-Caniego}, {M{\"u}ller}, {Nandra}, {Panuzzo}, {Papageorgiou},
  {Patanchon}, {Peacock}, {Pearson}, {Phillipps}, {Pohlen}, {Popescu},
  {Rawlings}, {Rigby}, {Rigopoulou}, {Robotham}, {Rodighiero}, {Sansom},
  {Schulz}, {Scott}, {Smith}, {Sibthorpe}, {Smail}, {Stevens}, {Sutherland},
  {Takeuchi}, {Tedds}, {Temi}, {Tuffs}, {Trichas}, {Vaccari}, {Valtchanov},
  {van der Werf}, {Verma}, {Vieria}, {Vlahakis}, \&
  {White}}]{2010PASP..122..499E}
{Eales}, S., {Dunne}, L., {Clements}, D., {et~al.} 2010, \pasp, 122, 499

\bibitem[{{Ferrarese} {et~al.}(2012){Ferrarese}, {C{\^o}t{\'e}}, {Cuillandre},
  {Gwyn}, {Peng}, {MacArthur}, {Duc}, {Boselli}, {Mei}, {Erben}, {McConnachie},
  {Durrell}, {Mihos}, {Jord{\'a}n}, {Lan{\c c}on}, {Puzia}, {Emsellem},
  {Balogh}, {Blakeslee}, {van Waerbeke}, {Gavazzi}, {Vollmer}, {Kavelaars},
  {Woods}, {Ball}, {Boissier}, {Courteau}, {Ferriere}, {Gavazzi},
  {Hildebrandt}, {Hudelot}, {Huertas-Company}, {Liu}, {McLaughlin}, {Mellier},
  {Milkeraitis}, {Schade}, {Balkowski}, {Bournaud}, {Carlberg}, {Chapman},
  {Hoekstra}, {Peng}, {Sawicki}, {Simard}, {Taylor}, {Tully}, {van Driel},
  {Wilson}, {Burdullis}, {Mahoney}, \& {Manset}}]{2012ApJS..200....4F}
{Ferrarese}, L., {C{\^o}t{\'e}}, P., {Cuillandre}, J.-C., {et~al.} 2012, \apjs,
  200, 4

\bibitem[{{Ferrarese} {et~al.}(2006){Ferrarese}, {C{\^o}t{\'e}}, {Jord{\'a}n},
  {Peng}, {Blakeslee}, {Piatek}, {Mei}, {Merritt}, {Milosavljevi{\'c}},
  {Tonry}, \& {West}}]{2006ApJS..164..334F}
{Ferrarese}, L., {C{\^o}t{\'e}}, P., {Jord{\'a}n}, A., {et~al.} 2006, \apjs,
  164, 334

\bibitem[{{Fry} \& {Gaztanaga}(1993)}]{1993ApJ...413..447F}
{Fry}, J.~N. \& {Gaztanaga}, E. 1993, \apj, 413, 447

\bibitem[{{Fu} {et~al.}(2012){Fu}, {Jullo}, {Cooray}, {Bussmann}, {Ivison},
  {P{\'e}rez-Fournon}, {Djorgovski}, {Scoville}, {Yan}, {Riechers}, {Aguirre},
  {Auld}, {Baes}, {Baker}, {Bradford}, {Cava}, {Clements}, {Dannerbauer},
  {Dariush}, {De Zotti}, {Dole}, {Dunne}, {Dye}, {Eales}, {Frayer}, {Gavazzi},
  {Gurwell}, {Harris}, {Herranz}, {Hopwood}, {Hoyos}, {Ibar}, {Jarvis}, {Kim},
  {Leeuw}, {Lupu}, {Maddox}, {Mart{\'{\i}}nez-Navajas}, {Micha{\l}owski},
  {Negrello}, {Omont}, {Rosenman}, {Scott}, {Serjeant}, {Smail}, {Swinbank},
  {Valiante}, {Verma}, {Vieira}, {Wardlow}, \& {van der
  Werf}}]{2012ApJ...753..134F}
{Fu}, H., {Jullo}, E., {Cooray}, A., {et~al.} 2012, \apj, 753, 134

\bibitem[{{Galametz} {et~al.}(2009){Galametz}, {Madden}, {Galliano}, {Hony},
  {Schuller}, {Beelen}, {Bendo}, {Sauvage}, {Lundgren}, \&
  {Billot}}]{2009A&A...508..645G}
{Galametz}, M., {Madden}, S., {Galliano}, F., {et~al.} 2009, \aap, 508, 645

\bibitem[{{Galametz} {et~al.}(2011){Galametz}, {Madden}, {Galliano}, {Hony},
  {Bendo}, \& {Sauvage}}]{2011A&A...532A..56G}
{Galametz}, M., {Madden}, S.~C., {Galliano}, F., {et~al.} 2011, \aap, 532, A56

\bibitem[{{Galliano} {et~al.}(2011){Galliano}, {Hony}, {Bernard}, {Bot},
  {Madden}, {Roman-Duval}, {Galametz}, {Li}, {Meixner}, {Engelbracht},
  {Lebouteiller}, {Misselt}, {Montiel}, {Panuzzo}, {Reach}, \&
  {Skibba}}]{2011A&A...536A..88G}
{Galliano}, F., {Hony}, S., {Bernard}, J.-P., {et~al.} 2011, \aap, 536, A88

\bibitem[{{Galliano} {et~al.}(2005){Galliano}, {Madden}, {Jones}, {Wilson}, \&
  {Bernard}}]{2005A&A...434..867G}
{Galliano}, F., {Madden}, S.~C., {Jones}, A.~P., {Wilson}, C.~D., \& {Bernard},
  J.-P. 2005, \aap, 434, 867

\bibitem[{{Galliano} {et~al.}(2003){Galliano}, {Madden}, {Jones}, {Wilson},
  {Bernard}, \& {Le Peintre}}]{2003A&A...407..159G}
{Galliano}, F., {Madden}, S.~C., {Jones}, A.~P., {et~al.} 2003, \aap, 407, 159

\bibitem[{{Gavazzi} {et~al.}(2003){Gavazzi}, {Boselli}, {Donati}, {Franzetti},
  \& {Scodeggio}}]{2003A&A...400..451G}
{Gavazzi}, G., {Boselli}, A., {Donati}, A., {Franzetti}, P., \& {Scodeggio}, M.
  2003, \aap, 400, 451

\bibitem[{{Gonz{\'a}lez-Nuevo} {et~al.}(2012){Gonz{\'a}lez-Nuevo}, {Lapi},
  {Fleuren}, {Bressan}, {Danese}, {De Zotti}, {Negrello}, {Cai}, {Fan},
  {Sutherland}, {Baes}, {Baker}, {Clements}, {Cooray}, {Dannerbauer}, {Dunne},
  {Dye}, {Eales}, {Frayer}, {Harris}, {Ivison}, {Jarvis}, {Micha{\l}owski},
  {L{\'o}pez-Caniego}, {Rodighiero}, {Rowlands}, {Serjeant}, {Scott}, {van der
  Werf}, {Auld}, {Buttiglione}, {Cava}, {Dariush}, {Fritz}, {Hopwood}, {Ibar},
  {Maddox}, {Pascale}, {Pohlen}, {Rigby}, {Smith}, \&
  {Temi}}]{2012ApJ...749...65G}
{Gonz{\'a}lez-Nuevo}, J., {Lapi}, A., {Fleuren}, S., {et~al.} 2012, \apj, 749,
  65

\bibitem[{{Griffin} {et~al.}(2010){Griffin}, {Abergel}, {Abreu}, {Ade},
  {Andr{\'e}}, {Augueres}, {Babbedge}, {Bae}, {Baillie}, {Baluteau}, {Barlow},
  {Bendo}, {Benielli}, {Bock}, {Bonhomme}, {Brisbin}, {Brockley-Blatt},
  {Caldwell}, {Cara}, {Castro-Rodriguez}, {Cerulli}, {Chanial}, {Chen},
  {Clark}, {Clements}, {Clerc}, {Coker}, {Communal}, {Conversi}, {Cox},
  {Crumb}, {Cunningham}, {Daly}, {Davis}, {de Antoni}, {Delderfield}, {Devin},
  {di Giorgio}, {Didschuns}, {Dohlen}, {Donati}, {Dowell}, {Dowell}, {Duband},
  {Dumaye}, {Emery}, {Ferlet}, {Ferrand}, {Fontignie}, {Fox}, {Franceschini},
  {Frerking}, {Fulton}, {Garcia}, {Gastaud}, {Gear}, {Glenn}, {Goizel},
  {Griffin}, {Grundy}, {Guest}, {Guillemet}, {Hargrave}, {Harwit}, {Hastings},
  {Hatziminaoglou}, {Herman}, {Hinde}, {Hristov}, {Huang}, {Imhof}, {Isaak},
  {Israelsson}, {Ivison}, {Jennings}, {Kiernan}, {King}, {Lange}, {Latter},
  {Laurent}, {Laurent}, {Leeks}, {Lellouch}, {Levenson}, {Li}, {Li},
  {Lilienthal}, {Lim}, {Liu}, {Lu}, {Madden}, {Mainetti}, {Marliani}, {McKay},
  {Mercier}, {Molinari}, {Morris}, {Moseley}, {Mulder}, {Mur}, {Naylor},
  {Nguyen}, {O'Halloran}, {Oliver}, {Olofsson}, {Olofsson}, {Orfei}, {Page},
  {Pain}, {Panuzzo}, {Papageorgiou}, {Parks}, {Parr-Burman}, {Pearce},
  {Pearson}, {P{\'e}rez-Fournon}, {Pinsard}, {Pisano}, {Podosek}, {Pohlen},
  {Polehampton}, {Pouliquen}, {Rigopoulou}, {Rizzo}, {Roseboom}, {Roussel},
  {Rowan-Robinson}, {Rownd}, {Saraceno}, {Sauvage}, {Savage}, {Savini},
  {Sawyer}, {Scharmberg}, {Schmitt}, {Schneider}, {Schulz}, {Schwartz},
  {Shafer}, {Shupe}, {Sibthorpe}, {Sidher}, {Smith}, {Smith}, {Smith},
  {Spencer}, {Stobie}, {Sudiwala}, {Sukhatme}, {Surace}, {Stevens}, {Swinyard},
  {Trichas}, {Tourette}, {Triou}, {Tseng}, {Tucker}, {Turner}, {Vaccari},
  {Valtchanov}, {Vigroux}, {Virique}, {Voellmer}, {Walker}, {Ward}, {Waskett},
  {Weilert}, {Wesson}, {White}, {Whitehouse}, {Wilson}, {Winter}, {Woodcraft},
  {Wright}, {Xu}, {Zavagno}, {Zemcov}, {Zhang}, \&
  {Zonca}}]{2010A&A...518L...3G}
{Griffin}, M.~J., {Abergel}, A., {Abreu}, A., {et~al.} 2010, \aap, 518, L3

\bibitem[{{Grossi} {et~al.}(2010){Grossi}, {Hunt}, {Madden}, {Vlahakis},
  {Bomans}, {Baes}, {Bendo}, {Bianchi}, {Boselli}, {Clemens}, {Corbelli},
  {Cortese}, {Dariush}, {Davies}, {De Looze}, {di Serego Alighieri}, {Fadda},
  {Fritz}, {Garcia-Appadoo}, {Gavazzi}, {Giovanardi}, {Hughes}, {Jones},
  {Pierini}, {Pohlen}, {Sabatini}, {Smith}, {Verstappen}, {Xilouris}, \&
  {Zibetti}}]{2010A&A...518L..52G}
{Grossi}, M., {Hunt}, L.~K., {Madden}, S., {et~al.} 2010, \aap, 518, L52

\bibitem[{{Harris} {et~al.}(2012){Harris}, {Baker}, {Frayer}, {Smail},
  {Swinbank}, {Riechers}, {van der Werf}, {Auld}, {Baes}, {Bussmann},
  {Buttiglione}, {Cava}, {Clements}, {Cooray}, {Dannerbauer}, {Dariush}, {De
  Zotti}, {Dunne}, {Dye}, {Eales}, {Fritz}, {Gonz{\'a}lez-Nuevo}, {Hopwood},
  {Ibar}, {Ivison}, {Jarvis}, {Maddox}, {Negrello}, {Rigby}, {Smith}, {Temi},
  \& {Wardlow}}]{2012ApJ...752..152H}
{Harris}, A.~I., {Baker}, A.~J., {Frayer}, D.~T., {et~al.} 2012, \apj, 752, 152

\bibitem[{{Hartogh} {et~al.}(2011){Hartogh}, {Lis}, {Bockel{\'e}e-Morvan}, {de
  Val-Borro}, {Biver}, {K{\"u}ppers}, {Emprechtinger}, {Bergin}, {Crovisier},
  {Rengel}, {Moreno}, {Szutowicz}, \& {Blake}}]{2011Natur.478..218H}
{Hartogh}, P., {Lis}, D.~C., {Bockel{\'e}e-Morvan}, D., {et~al.} 2011, \nat,
  478, 218

\bibitem[{{Herranz} {et~al.}(2013){Herranz}, {Gonz{\'a}lez-Nuevo}, {Clements},
  {De Zotti}, {Lopez-Caniego}, {Lapi}, {Rodighiero}, {Danese}, {Fu}, {Cooray},
  {Baes}, {Bendo}, {Bonavera}, {Carrera}, {Dole}, {Eales}, {Ivison}, {Jarvis},
  {Lagache}, {Massardi}, {Micha{\l}owski}, {Negrello}, {Rigby}, {Scott},
  {Valiante}, {Valtchanov}, {Van der Werf}, {Auld}, {Buttiglione}, {Dariush},
  {Dunne}, {Hopwood}, {Hoyos}, {Ibar}, \& {Maddox}}]{2013A&A...549A..31H}
{Herranz}, D., {Gonz{\'a}lez-Nuevo}, J., {Clements}, D.~L., {et~al.} 2013,
  \aap, 549, A31

\bibitem[{{Holland} {et~al.}(2006){Holland}, {MacIntosh}, {Fairley}, {Kelly},
  {Montgomery}, {Gostick}, {Atad-Ettedgui}, {Ellis}, {Robson}, {Hollister},
  {Woodcraft}, {Ade}, {Walker}, {Irwin}, {Hilton}, {Duncan}, {Reintsema},
  {Walton}, {Parkes}, {Dunare}, {Fich}, {Kycia}, {Halpern}, {Scott}, {Gibb},
  {Molnar}, {Chapin}, {Bintley}, {Craig}, {Chylek}, {Jenness}, {Economou}, \&
  {Davis}}]{2006SPIE.6275E..45H}
{Holland}, W., {MacIntosh}, M., {Fairley}, A., {et~al.} 2006, in Society of
  Photo-Optical Instrumentation Engineers (SPIE) Conference Series, Vol. 6275,
  Society of Photo-Optical Instrumentation Engineers (SPIE) Conference Series

\bibitem[{{Holland} {et~al.}(1999){Holland}, {Robson}, {Gear}, {Cunningham},
  {Lightfoot}, {Jenness}, {Ivison}, {Stevens}, {Ade}, {Griffin}, {Duncan},
  {Murphy}, \& {Naylor}}]{1999MNRAS.303..659H}
{Holland}, W.~S., {Robson}, E.~I., {Gear}, W.~K., {et~al.} 1999, \mnras, 303,
  659

\bibitem[{{Israel} {et~al.}(2010){Israel}, {Wall}, {Raban}, {Reach}, {Bot},
  {Oonk}, {Ysard}, \& {Bernard}}]{2010A&A...519A..67I}
{Israel}, F.~P., {Wall}, W.~F., {Raban}, D., {et~al.} 2010, \aap, 519, A67

\bibitem[{{Kenney} {et~al.}(1990){Kenney}, {Young}, {Hasegawa}, \&
  {Nakai}}]{1990ApJ...353..460K}
{Kenney}, J.~D.~P., {Young}, J.~S., {Hasegawa}, T., \& {Nakai}, N. 1990, \apj,
  353, 460

\bibitem[{{Kennicutt} {et~al.}(2011){Kennicutt}, {Calzetti}, {Aniano},
  {Appleton}, {Armus}, {Beir{\~a}o}, {Bolatto}, {Brandl}, {Crocker}, {Croxall},
  {Dale}, {Meyer}, {Draine}, {Engelbracht}, {Galametz}, {Gordon}, {Groves},
  {Hao}, {Helou}, {Hinz}, {Hunt}, {Johnson}, {Koda}, {Krause}, {Leroy}, {Li},
  {Meidt}, {Montiel}, {Murphy}, {Rahman}, {Rix}, {Roussel}, {Sandstrom},
  {Sauvage}, {Schinnerer}, {Skibba}, {Smith}, {Srinivasan}, {Vigroux},
  {Walter}, {Wilson}, {Wolfire}, \& {Zibetti}}]{2011PASP..123.1347K}
{Kennicutt}, R.~C., {Calzetti}, D., {Aniano}, G., {et~al.} 2011, \pasp, 123,
  1347

\bibitem[{{Kessler} {et~al.}(1996){Kessler}, {Steinz}, {Anderegg}, {Clavel},
  {Drechsel}, {Estaria}, {Faelker}, {Riedinger}, {Robson}, {Taylor}, \&
  {Xim{\'e}nez de Ferr{\'a}n}}]{1996A&A...315L..27K}
{Kessler}, M.~F., {Steinz}, J.~A., {Anderegg}, M.~E., {et~al.} 1996, \aap, 315,
  L27

\bibitem[{{Lamarre} {et~al.}(2010){Lamarre}, {Puget}, {Ade}, {Bouchet},
  {Guyot}, {Lange}, {Pajot}, {Arondel}, {Benabed}, {Beney}, {Beno{\^i}t},
  {Bernard}, {Bhatia}, {Blanc}, {Bock}, {Br{\'e}elle}, {Bradshaw}, {Camus},
  {Catalano}, {Charra}, {Charra}, {Church}, {Couchot}, {Coulais}, {Crill},
  {Crook}, {Dassas}, {de Bernardis}, {Delabrouille}, {de Marcillac}, {Delouis},
  {D{\'e}sert}, {Dumesnil}, {Dupac}, {Efstathiou}, {Eng}, {Evesque},
  {Fourmond}, {Ganga}, {Giard}, {Gispert}, {Guglielmi}, {Haissinski},
  {Henrot-Versill{\'e}}, {Hivon}, {Holmes}, {Jones}, {Koch}, {Lagard{\`e}re},
  {Lami}, {Land{\'e}}, {Leriche}, {Leroy}, {Longval},
  {Mac{\'{\i}}as-P{\'e}rez}, {Maciaszek}, {Maffei}, {Mansoux}, {Marty}, {Masi},
  {Mercier}, {Miville-Desch{\^e}nes}, {Moneti}, {Montier}, {Murphy},
  {Narbonne}, {Nexon}, {Paine}, {Pahn}, {Perdereau}, {Piacentini}, {Piat},
  {Plaszczynski}, {Pointecouteau}, {Pons}, {Ponthieu}, {Prunet}, {Rambaud},
  {Recouvreur}, {Renault}, {Ristorcelli}, {Rosset}, {Santos}, {Savini},
  {Serra}, {Stassi}, {Sudiwala}, {Sygnet}, {Tauber}, {Torre}, {Tristram},
  {Vibert}, {Woodcraft}, {Yurchenko}, \& {Yvon}}]{2010A&A...520A...9L}
{Lamarre}, J.-M., {Puget}, J.-L., {Ade}, P.~A.~R., {et~al.} 2010, \aap, 520, A9

\bibitem[{{Lapi} {et~al.}(2011){Lapi}, {Gonz{\'a}lez-Nuevo}, {Fan}, {Bressan},
  {De Zotti}, {Danese}, {Negrello}, {Dunne}, {Eales}, {Maddox}, {Auld}, {Baes},
  {Bonfield}, {Buttiglione}, {Cava}, {Clements}, {Cooray}, {Dariush}, {Dye},
  {Fritz}, {Herranz}, {Hopwood}, {Ibar}, {Ivison}, {Jarvis}, {Kaviraj},
  {L{\'o}pez-Caniego}, {Massardi}, {Micha{\l}owski}, {Pascale}, {Pohlen},
  {Rigby}, {Rodighiero}, {Serjeant}, {Smith}, {Temi}, {Wardlow}, \& {van der
  Werf}}]{2011ApJ...742...24L}
{Lapi}, A., {Gonz{\'a}lez-Nuevo}, J., {Fan}, L., {et~al.} 2011, \apj, 742, 24

\bibitem[{{Lapi} {et~al.}(2012){Lapi}, {Negrello}, {Gonz{\'a}lez-Nuevo}, {Cai},
  {De Zotti}, \& {Danese}}]{2012ApJ...755...46L}
{Lapi}, A., {Negrello}, M., {Gonz{\'a}lez-Nuevo}, J., {et~al.} 2012, \apj, 755,
  46

\bibitem[{{Lawrence} {et~al.}(2007){Lawrence}, {Warren}, {Almaini}, {Edge},
  {Hambly}, {Jameson}, {Lucas}, {Casali}, {Adamson}, {Dye}, {Emerson},
  {Foucaud}, {Hewett}, {Hirst}, {Hodgkin}, {Irwin}, {Lodieu}, {McMahon},
  {Simpson}, {Smail}, {Mortlock}, \& {Folger}}]{2007MNRAS.379.1599L}
{Lawrence}, A., {Warren}, S.~J., {Almaini}, O., {et~al.} 2007, \mnras, 379,
  1599

\bibitem[{{Maddox} {et~al.}(2010){Maddox}, {Dunne}, {Rigby}, {Eales}, {Cooray},
  {Scott}, {Peacock}, {Negrello}, {Smith}, {Benford}, {Amblard}, {Auld},
  {Baes}, {Bonfield}, {Burgarella}, {Buttiglione}, {Cava}, {Clements},
  {Dariush}, {de Zotti}, {Dye}, {Frayer}, {Fritz}, {Gonzalez-Nuevo}, {Herranz},
  {Ibar}, {Ivison}, {Jarvis}, {Lagache}, {Leeuw}, {Lopez-Caniego}, {Pascale},
  {Pohlen}, {Rodighiero}, {Samui}, {Serjeant}, {Temi}, {Thompson}, \&
  {Verma}}]{2010A&A...518L..11M}
{Maddox}, S.~J., {Dunne}, L., {Rigby}, E., {et~al.} 2010, \aap, 518, L11

\bibitem[{{Magrini} {et~al.}(2011){Magrini}, {Bianchi}, {Corbelli}, {Cortese},
  {Hunt}, {Smith}, {Vlahakis}, {Davies}, {Bendo}, {Baes}, {Boselli}, {Clemens},
  {Casasola}, {De Looze}, {Fritz}, {Giovanardi}, {Grossi}, {Hughes}, {Madden},
  {Pappalardo}, {Pohlen}, {di Serego Alighieri}, \&
  {Verstappen}}]{2011A&A...535A..13M}
{Magrini}, L., {Bianchi}, S., {Corbelli}, E., {et~al.} 2011, \aap, 535, A13

\bibitem[{{Markwardt}(2009)}]{2009ASPC..411..251M}
{Markwardt}, C.~B. 2009, in Astronomical Society of the Pacific Conference
  Series, Vol. 411, Astronomical Data Analysis Software and Systems XVIII, ed.
  D.~A. {Bohlender}, D.~{Durand}, \& P.~{Dowler}, 251

\bibitem[{{Mei} {et~al.}(2007){Mei}, {Blakeslee}, {C{\^o}t{\'e}}, {Tonry},
  {West}, {Ferrarese}, {Jord{\'a}n}, {Peng}, {Anthony}, \&
  {Merritt}}]{2007ApJ...655..144M}
{Mei}, S., {Blakeslee}, J.~P., {C{\^o}t{\'e}}, P., {et~al.} 2007, \apj, 655,
  144

\bibitem[{{Mihos} {et~al.}(2005){Mihos}, {Harding}, {Feldmeier}, \&
  {Morrison}}]{2005ApJ...631L..41M}
{Mihos}, J.~C., {Harding}, P., {Feldmeier}, J., \& {Morrison}, H. 2005, \apjl,
  631, L41

\bibitem[{{Mitra} {et~al.}(2011){Mitra}, {Rocha}, {G{\'o}rski}, {Huffenberger},
  {Eriksen}, {Ashdown}, \& {Lawrence}}]{2011ApJS..193....5M}
{Mitra}, S., {Rocha}, G., {G{\'o}rski}, K.~M., {et~al.} 2011, \apjs, 193, 5

\bibitem[{{Mould} {et~al.}(2000){Mould}, {Huchra}, {Freedman}, {Kennicutt},
  {Ferrarese}, {Ford}, {Gibson}, {Graham}, {Hughes}, {Illingworth}, {Kelson},
  {Macri}, {Madore}, {Sakai}, {Sebo}, {Silbermann}, \&
  {Stetson}}]{2000ApJ...529..786M}
{Mould}, J.~R., {Huchra}, J.~P., {Freedman}, W.~L., {et~al.} 2000, \apj, 529,
  786

\bibitem[{{M{\"u}ller} {et~al.}(2010){M{\"u}ller}, {Lellouch}, {Stansberry},
  {Kiss}, {Santos-Sanz}, {Vilenius}, {Protopapa}, {Moreno}, {Mueller},
  {Delsanti}, {Duffard}, {Fornasier}, {Groussin}, {Harris}, {Henry}, {Horner},
  {Lacerda}, {Lim}, {Mommert}, {Ortiz}, {Rengel}, {Thirouin}, {Trilling},
  {Barucci}, {Crovisier}, {Doressoundiram}, {Dotto}, {Guti{\'e}rrez},
  {Hainaut}, {Hartogh}, {Hestroffer}, {Kidger}, {Lara}, {Swinyard}, \&
  {Thomas}}]{2010A&A...518L.146M}
{M{\"u}ller}, T.~G., {Lellouch}, E., {Stansberry}, J., {et~al.} 2010, \aap,
  518, L146

\bibitem[{{Murakami} {et~al.}(2007){Murakami}, {Baba}, {Barthel}, {Clements},
  {Cohen}, {Doi}, {Enya}, {Figueredo}, {Fujishiro}, {Fujiwara}, {Fujiwara},
  {Garcia-Lario}, {Goto}, {Hasegawa}, {Hibi}, {Hirao}, {Hiromoto}, {Hong},
  {Imai}, {Ishigaki}, {Ishiguro}, {Ishihara}, {Ita}, {Jeong}, {Jeong},
  {Kaneda}, {Kataza}, {Kawada}, {Kawai}, {Kawamura}, {Kessler}, {Kester},
  {Kii}, {Kim}, {Kim}, {Kobayashi}, {Koo}, {Kwon}, {Lee}, {Lorente}, {Makiuti},
  {Matsuhara}, {Matsumoto}, {Matsuo}, {Matsuura}, {M{\"u}ller}, {Murakami},
  {Nagata}, {Nakagawa}, {Naoi}, {Narita}, {Noda}, {Oh}, {Ohnishi}, {Ohyama},
  {Okada}, {Okuda}, {Oliver}, {Onaka}, {Ootsubo}, {Oyabu}, {Pak}, {Park},
  {Pearson}, {Rowan-Robinson}, {Saito}, {Sakon}, {Salama}, {Sato}, {Savage},
  {Serjeant}, {Shibai}, {Shirahata}, {Sohn}, {Suzuki}, {Takagi}, {Takahashi},
  {Tanab{\'e}}, {Takeuchi}, {Takita}, {Thomson}, {Uemizu}, {Ueno}, {Usui},
  {Verdugo}, {Wada}, {Wang}, {Watabe}, {Watarai}, {White}, {Yamamura},
  {Yamauchi}, \& {Yasuda}}]{2007PASJ...59S.369M}
{Murakami}, H., {Baba}, H., {Barthel}, P., {et~al.} 2007, \pasj, 59, 369

\bibitem[{{Negrello} {et~al.}(2013){Negrello}, {Clemens}, {Gonzalez-Nuevo}, {De
  Zotti}, {Bonavera}, {Cosco}, {Guarese}, {Boaretto}, {Serjeant}, {Toffolatti},
  {Lapi}, {Bethermin}, {Castex}, {Clements}, {Delabrouille}, {Dole},
  {Franceschini}, {Mandolesi}, {Marchetti}, {Partridge}, \&
  {Sajina}}]{2013MNRAS.429.1309N}
{Negrello}, M., {Clemens}, M., {Gonzalez-Nuevo}, J., {et~al.} 2013, \mnras,
  429, 1309

\bibitem[{{Negrello} {et~al.}(2005){Negrello}, {Gonz{\'a}lez-Nuevo},
  {Magliocchetti}, {Moscardini}, {De Zotti}, {Toffolatti}, \&
  {Danese}}]{2005MNRAS.358..869N}
{Negrello}, M., {Gonz{\'a}lez-Nuevo}, J., {Magliocchetti}, M., {et~al.} 2005,
  \mnras, 358, 869

\bibitem[{{Negrello} {et~al.}(2010){Negrello}, {Hopwood}, {De Zotti}, {Cooray},
  {Verma}, {Bock}, {Frayer}, {Gurwell}, {Omont}, {Neri}, {Dannerbauer},
  {Leeuw}, {Barton}, {Cooke}, {Kim}, {da Cunha}, {Rodighiero}, {Cox},
  {Bonfield}, {Jarvis}, {Serjeant}, {Ivison}, {Dye}, {Aretxaga}, {Hughes},
  {Ibar}, {Bertoldi}, {Valtchanov}, {Eales}, {Dunne}, {Driver}, {Auld},
  {Buttiglione}, {Cava}, {Grady}, {Clements}, {Dariush}, {Fritz}, {Hill},
  {Hornbeck}, {Kelvin}, {Lagache}, {Lopez-Caniego}, {Gonzalez-Nuevo}, {Maddox},
  {Pascale}, {Pohlen}, {Rigby}, {Robotham}, {Simpson}, {Smith}, {Temi},
  {Thompson}, {Woodgate}, {York}, {Aguirre}, {Beelen}, {Blain}, {Baker},
  {Birkinshaw}, {Blundell}, {Bradford}, {Burgarella}, {Danese}, {Dunlop},
  {Fleuren}, {Glenn}, {Harris}, {Kamenetzky}, {Lupu}, {Maddalena}, {Madore},
  {Maloney}, {Matsuhara}, {Michaowski}, {Murphy}, {Naylor}, {Nguyen},
  {Popescu}, {Rawlings}, {Rigopoulou}, {Scott}, {Scott}, {Seibert}, {Smail},
  {Tuffs}, {Vieira}, {van der Werf}, \& {Zmuidzinas}}]{2010Sci...330..800N}
{Negrello}, M., {Hopwood}, R., {De Zotti}, G., {et~al.} 2010, Science, 330, 800

\bibitem[{{Negrello} {et~al.}(2007){Negrello}, {Perrotta},
  {Gonz{\'a}lez-Nuevo}, {Silva}, {de Zotti}, {Granato}, {Baccigalupi}, \&
  {Danese}}]{2007MNRAS.377.1557N}
{Negrello}, M., {Perrotta}, F., {Gonz{\'a}lez-Nuevo}, J., {et~al.} 2007,
  \mnras, 377, 1557

\bibitem[{{Neugebauer} {et~al.}(1984){Neugebauer}, {Habing}, {van Duinen},
  {Aumann}, {Baud}, {Beichman}, {Beintema}, {Boggess}, {Clegg}, {de Jong},
  {Emerson}, {Gautier}, {Gillett}, {Harris}, {Hauser}, {Houck}, {Jennings},
  {Low}, {Marsden}, {Miley}, {Olnon}, {Pottasch}, {Raimond}, {Rowan-Robinson},
  {Soifer}, {Walker}, {Wesselius}, \& {Young}}]{1984ApJ...278L...1N}
{Neugebauer}, G., {Habing}, H.~J., {van Duinen}, R., {et~al.} 1984, \apjl, 278,
  L1

\bibitem[{{Oliver} {et~al.}(2012){Oliver}, {Bock}, {Altieri}, {Amblard},
  {Arumugam}, {Aussel}, {Babbedge}, {Beelen}, {B{\'e}thermin}, {Blain},
  {Boselli}, {Bridge}, {Brisbin}, {Buat}, {Burgarella},
  {Castro-Rodr{\'{\i}}guez}, {Cava}, {Chanial}, {Cirasuolo}, {Clements},
  {Conley}, {Conversi}, {Cooray}, {Dowell}, {Dubois}, {Dwek}, {Dye}, {Eales},
  {Elbaz}, {Farrah}, {Feltre}, {Ferrero}, {Fiolet}, {Fox}, {Franceschini},
  {Gear}, {Giovannoli}, {Glenn}, {Gong}, {Gonz{\'a}lez Solares}, {Griffin},
  {Halpern}, {Harwit}, {Hatziminaoglou}, {Heinis}, {Hurley}, {Hwang}, {Hyde},
  {Ibar}, {Ilbert}, {Isaak}, {Ivison}, {Lagache}, {Le Floc'h}, {Levenson},
  {Faro}, {Lu}, {Madden}, {Maffei}, {Magdis}, {Mainetti}, {Marchetti},
  {Marsden}, {Marshall}, {Mortier}, {Nguyen}, {O'Halloran}, {Omont}, {Page},
  {Panuzzo}, {Papageorgiou}, {Patel}, {Pearson}, {P{\'e}rez-Fournon}, {Pohlen},
  {Rawlings}, {Raymond}, {Rigopoulou}, {Riguccini}, {Rizzo}, {Rodighiero},
  {Roseboom}, {Rowan-Robinson}, {S{\'a}nchez Portal}, {Schulz}, {Scott},
  {Seymour}, {Shupe}, {Smith}, {Stevens}, {Symeonidis}, {Trichas}, {Tugwell},
  {Vaccari}, {Valtchanov}, {Vieira}, {Viero}, {Vigroux}, {Wang}, {Ward},
  {Wardlow}, {Wright}, {Xu}, \& {Zemcov}}]{2012MNRAS.424.1614O}
{Oliver}, S.~J., {Bock}, J., {Altieri}, B., {et~al.} 2012, \mnras, 424, 1614

\bibitem[{{Pascale} {et~al.}(2008){Pascale}, {Ade}, {Bock}, {Chapin}, {Chung},
  {Devlin}, {Dicker}, {Griffin}, {Gundersen}, {Halpern}, {Hargrave}, {Hughes},
  {Klein}, {MacTavish}, {Marsden}, {Martin}, {Martin}, {Mauskopf},
  {Netterfield}, {Olmi}, {Patanchon}, {Rex}, {Scott}, {Semisch}, {Thomas},
  {Truch}, {Tucker}, {Tucker}, {Viero}, \& {Wiebe}}]{2008ApJ...681..400P}
{Pascale}, E., {Ade}, P.~A.~R., {Bock}, J.~J., {et~al.} 2008, \apj, 681, 400

\bibitem[{{Pilbratt} {et~al.}(2010){Pilbratt}, {Riedinger}, {Passvogel},
  {Crone}, {Doyle}, {Gageur}, {Heras}, {Jewell}, {Metcalfe}, {Ott}, \&
  {Schmidt}}]{2010A&A...518L...1P}
{Pilbratt}, G.~L., {Riedinger}, J.~R., {Passvogel}, T., {et~al.} 2010, \aap,
  518, L1

\bibitem[{{Planck Collaboration}(2011{\natexlab{a}})}]{2011A&A...536A...1P}
{Planck Collaboration}. 2011{\natexlab{a}}, \aap, 536, A1

\bibitem[{{Planck Collaboration}(2011{\natexlab{b}})}]{2011A&A...536A...7P}
{Planck Collaboration}. 2011{\natexlab{b}}, \aap, 536, A7

\bibitem[{{Planck Collaboration}(2011{\natexlab{c}})}]{2011A&A...536A..19P}
{Planck Collaboration}. 2011{\natexlab{c}}, \aap, 536, A19

\bibitem[{{Planck Collaboration}(2011{\natexlab{d}})}]{2011A&A...536A..15P}
{Planck Collaboration}. 2011{\natexlab{d}}, \aap, 536, A15

\bibitem[{{Planck Collaboration}(2011{\natexlab{e}})}]{2011A&A...536A..16P}
{Planck Collaboration}. 2011{\natexlab{e}}, \aap, 536, A16

\bibitem[{{Planck Collaboration}(2011{\natexlab{f}})}]{2011A&A...536A..17P}
{Planck Collaboration}. 2011{\natexlab{f}}, \aap, 536, A17

\bibitem[{{Planck Collaboration}(2011{\natexlab{g}})}]{2011A&A...536A..21P}
{Planck Collaboration}. 2011{\natexlab{g}}, \aap, 536, A21

\bibitem[{{Planck Collaboration}(2011{\natexlab{h}})}]{2011A&A...536A..22P}
{Planck Collaboration}. 2011{\natexlab{h}}, \aap, 536, A22

\bibitem[{{Planck Collaboration}(2011{\natexlab{i}})}]{2011A&A...536A..25P}
{Planck Collaboration}. 2011{\natexlab{i}}, \aap, 536, A25

\bibitem[{{Planck Collaboration}(2013{\natexlab{a}})}]{2013arXiv1303.5062P}
{Planck Collaboration}. 2013{\natexlab{a}}, \aap, submitted, arXiv:1303.5062

\bibitem[{{Planck Collaboration}(2013{\natexlab{b}})}]{2013arXiv1303.5075P}
{Planck Collaboration}. 2013{\natexlab{b}}, \aap, submitted, arXiv:1303.5075

\bibitem[{{Planck Collaboration}(2013{\natexlab{c}})}]{2013arXiv1303.5076P}
{Planck Collaboration}. 2013{\natexlab{c}}, \aap, submitted, arXiv:1303.5076

\bibitem[{{Planck Collaboration}(2013{\natexlab{d}})}]{2013arXiv1303.5083P}
{Planck Collaboration}. 2013{\natexlab{d}}, \aap, submitted, arXiv:1303.5076

\bibitem[{{Planck Collaboration}(2013{\natexlab{e}})}]{2013arXiv1303.5088P}
{Planck Collaboration}. 2013{\natexlab{e}}, \aap, submitted, arXiv:1303.5088

\bibitem[{{Poglitsch} {et~al.}(2010){Poglitsch}, {Waelkens}, {Geis},
  {Feuchtgruber}, {Vandenbussche}, {Rodriguez}, {Krause}, {Renotte}, {van
  Hoof}, {Saraceno}, {Cepa}, {Kerschbaum}, {Agn{\`e}se}, {Ali}, {Altieri},
  {Andreani}, {Augueres}, {Balog}, {Barl}, {Bauer}, {Belbachir}, {Benedettini},
  {Billot}, {Boulade}, {Bischof}, {Blommaert}, {Callut}, {Cara}, {Cerulli},
  {Cesarsky}, {Contursi}, {Creten}, {De Meester}, {Doublier}, {Doumayrou},
  {Duband}, {Exter}, {Genzel}, {Gillis}, {Gr{\"o}zinger}, {Henning},
  {Herreros}, {Huygen}, {Inguscio}, {Jakob}, {Jamar}, {Jean}, {de Jong},
  {Katterloher}, {Kiss}, {Klaas}, {Lemke}, {Lutz}, {Madden}, {Marquet},
  {Martignac}, {Mazy}, {Merken}, {Montfort}, {Morbidelli}, {M{\"u}ller},
  {Nielbock}, {Okumura}, {Orfei}, {Ottensamer}, {Pezzuto}, {Popesso},
  {Putzeys}, {Regibo}, {Reveret}, {Royer}, {Sauvage}, {Schreiber}, {Stegmaier},
  {Schmitt}, {Schubert}, {Sturm}, {Thiel}, {Tofani}, {Vavrek}, {Wetzstein},
  {Wieprecht}, \& {Wiezorrek}}]{2010A&A...518L...2P}
{Poglitsch}, A., {Waelkens}, C., {Geis}, N., {et~al.} 2010, \aap, 518, L2

\bibitem[{{Rudick} {et~al.}(2010){Rudick}, {Mihos}, {Harding}, {Feldmeier},
  {Janowiecki}, \& {Morrison}}]{2010ApJ...720..569R}
{Rudick}, C.~S., {Mihos}, J.~C., {Harding}, P., {et~al.} 2010, \apj, 720, 569

\bibitem[{{Sandage} \& {Bedke}(1994)}]{1994cag..book.....S}
{Sandage}, A. \& {Bedke}, J. 1994, {The Carnegie Atlas of Galaxies} ({Carnegie
  Institution of Washington Publ.})

\bibitem[{{Sarzi} {et~al.}(2006){Sarzi}, {Falc{\'o}n-Barroso}, {Davies},
  {Bacon}, {Bureau}, {Cappellari}, {de Zeeuw}, {Emsellem}, {Fathi},
  {Krajnovi{\'c}}, {Kuntschner}, {McDermid}, \&
  {Peletier}}]{2006MNRAS.366.1151S}
{Sarzi}, M., {Falc{\'o}n-Barroso}, J., {Davies}, R.~L., {et~al.} 2006, \mnras,
  366, 1151

\bibitem[{{Shi} {et~al.}(2007){Shi}, {Rieke}, {Hines}, {Gordon}, \&
  {Egami}}]{2007ApJ...655..781S}
{Shi}, Y., {Rieke}, G.~H., {Hines}, D.~C., {Gordon}, K.~D., \& {Egami}, E.
  2007, \apj, 655, 781

\bibitem[{{Short} \& {Coles}(2011)}]{2011MNRAS.412L..93S}
{Short}, J. \& {Coles}, P. 2011, \mnras, 412, L93

\bibitem[{{Siringo} {et~al.}(2009){Siringo}, {Kreysa}, {Kov{\'a}cs},
  {Schuller}, {Wei{\ss}}, {Esch}, {Gem{\"u}nd}, {Jethava}, {Lundershausen},
  {Colin}, {G{\"u}sten}, {Menten}, {Beelen}, {Bertoldi}, {Beeman}, \&
  {Haller}}]{2009A&A...497..945S}
{Siringo}, G., {Kreysa}, E., {Kov{\'a}cs}, A., {et~al.} 2009, \aap, 497, 945

\bibitem[{{Smith} {et~al.}(2013){Smith}, {Hardcastle}, {Jarvis}, {Maddox},
  {Dunne}, {Bonfield}, {Eales}, {Serjeant}, {Thompson}, {Baes}, {Clements},
  {Cooray}, {De Zotti}, {Gonz{\`a}lez-Nuevo}, {Werf}, {Virdee}, {Bourne},
  {Dariush}, {Hopwood}, {Ibar}, \& {Valiante}}]{2013MNRAS.436.2435S}
{Smith}, D.~J.~B., {Hardcastle}, M.~J., {Jarvis}, M.~J., {et~al.} 2013, \mnras,
  436, 2435

\bibitem[{{Smith} {et~al.}(2010){Smith}, {Vlahakis}, {Baes}, {Bendo},
  {Bianchi}, {Bomans}, {Boselli}, {Clemens}, {Corbelli}, {Cortese}, {Dariush},
  {Davies}, {De Looze}, {di Serego Alighieri}, {Fadda}, {Fritz},
  {Garcia-Appadoo}, {Gavazzi}, {Giovanardi}, {Grossi}, {Hughes}, {Hunt},
  {Jones}, {Madden}, {Pierini}, {Pohlen}, {Sabatini}, {Verstappen}, {Xilouris},
  \& {Zibetti}}]{2010A&A...518L..51S}
{Smith}, M.~W.~L., {Vlahakis}, C., {Baes}, M., {et~al.} 2010, \aap, 518, L51

\bibitem[{{Springob} {et~al.}(2005){Springob}, {Haynes}, {Giovanelli}, \&
  {Kent}}]{2005ApJS..160..149S}
{Springob}, C.~M., {Haynes}, M.~P., {Giovanelli}, R., \& {Kent}, B.~R. 2005,
  \apjs, 160, 149

\bibitem[{{Springob} {et~al.}(2009){Springob}, {Masters}, {Haynes},
  {Giovanelli}, \& {Marinoni}}]{2009ApJS..182..474S}
{Springob}, C.~M., {Masters}, K.~L., {Haynes}, M.~P., {Giovanelli}, R., \&
  {Marinoni}, C. 2009, \apjs, 182, 474

\bibitem[{{Szapudi} {et~al.}(2001){Szapudi}, {Postman}, {Lauer}, \&
  {Oegerle}}]{2001ApJ...548..114S}
{Szapudi}, I., {Postman}, M., {Lauer}, T.~R., \& {Oegerle}, W. 2001, \apj, 548,
  114

\bibitem[{{Thacker} {et~al.}(2013){Thacker}, {Cooray}, {Smidt}, {De Bernardis},
  {Mitchell-Wynne}, {Amblard}, {Auld}, {Baes}, {Clements}, {Dariush}, {De
  Zotti}, {Dunne}, {Eales}, {Hopwood}, {Hoyos}, {Ibar}, {Jarvis}, {Maddox},
  {Micha{\l}owski}, {Pascale}, {Scott}, {Serjeant}, {Smith}, {Valiante}, \&
  {van der Werf}}]{2013ApJ...768...58T}
{Thacker}, C., {Cooray}, A., {Smidt}, J., {et~al.} 2013, \apj, 768, 58

\bibitem[{{van Kampen} {et~al.}(2012){van Kampen}, {Smith}, {Maddox},
  {Hopkins}, {Valtchanov}, {Peacock}, {Micha{\l}owski}, {Norberg}, {Eales},
  {Dunne}, {Liske}, {Baes}, {Scott}, {Rigby}, {Robotham}, {van der Werf},
  {Ibar}, {Jarvis}, {Loveday}, {Auld}, {Baldry}, {Bamford}, {Cameron}, {Croom},
  {Buttiglione}, {Cava}, {Cooray}, {Driver}, {Dunlop}, {Dariush}, {Fritz},
  {Ivison}, {Pascale}, {Pohlen}, {Rodighiero}, {Temi}, {Bonfield}, {Hill},
  {Jones}, {Kelvin}, {Parkinson}, {Prescott}, {Sharp}, {de Zotti}, {Serjeant},
  {Popescu}, \& {Tuffs}}]{2012MNRAS.426.3455V}
{van Kampen}, E., {Smith}, D.~J.~B., {Maddox}, S., {et~al.} 2012, \mnras, 426,
  3455

\bibitem[{{Werner} {et~al.}(2004){Werner}, {Roellig}, {Low}, {Rieke}, {Rieke},
  {Hoffmann}, {Young}, {Houck}, {Brandl}, {Fazio}, {Hora}, {Gehrz}, {Helou},
  {Soifer}, {Stauffer}, {Keene}, {Eisenhardt}, {Gallagher}, {Gautier}, {Irace},
  {Lawrence}, {Simmons}, {Van Cleve}, {Jura}, {Wright}, \&
  {Cruikshank}}]{2004ApJS..154....1W}
{Werner}, M.~W., {Roellig}, T.~L., {Low}, F.~J., {et~al.} 2004, \apjs, 154, 1

\bibitem[{{Williams} {et~al.}(2010){Williams}, {Bureau}, \&
  {Cappellari}}]{2010MNRAS.409.1330W}
{Williams}, M.~J., {Bureau}, M., \& {Cappellari}, M. 2010, \mnras, 409, 1330

\bibitem[{{Wright} {et~al.}(2010){Wright}, {Eisenhardt}, {Mainzer}, {Ressler},
  {Cutri}, {Jarrett}, {Kirkpatrick}, {Padgett}, {McMillan}, {Skrutskie},
  {Stanford}, {Cohen}, {Walker}, {Mather}, {Leisawitz}, {Gautier}, {McLean},
  {Benford}, {Lonsdale}, {Blain}, {Mendez}, {Irace}, {Duval}, {Liu}, {Royer},
  {Heinrichsen}, {Howard}, {Shannon}, {Kendall}, {Walsh}, {Larsen}, {Cardon},
  {Schick}, {Schwalm}, {Abid}, {Fabinsky}, {Naes}, \&
  {Tsai}}]{2010AJ....140.1868W}
{Wright}, E.~L., {Eisenhardt}, P.~R.~M., {Mainzer}, A.~K., {et~al.} 2010, \aj,
  140, 1868

\bibitem[{{Young} {et~al.}(2009){Young}, {Bendo}, \&
  {Lucero}}]{2009AJ....137.3053Y}
{Young}, L.~M., {Bendo}, G.~J., \& {Lucero}, D.~M. 2009, \aj, 137, 3053

\end{thebibliography}

\end{document}